\documentclass[11pt,psfig]{article}
\usepackage{graphicx}
\title{Statistical Error in Particle
Simulations\\ of Hydrodynamic Phenomena}
\author{
Nicolas~G.~Hadjiconstantinou\\
{\it Department of Mechanical Engineering}\\
{\it Massachusetts Institute of Technology},
{\it Cambridge, MA 02139}\\
\ \\
Alejandro~L. Garcia\thanks{Permanent address: Dept. of Physics,
San Jose State Univ., San Jose, CA 95192-0106}\\
{\it Center for Applied Scientific Computing}\\
{\it Lawrence Livermore National Laboratory},
{\it Livermore, CA 94551} \\ 
\ \\
Martin~ Z. ~Bazant\\
{\it Department of Mathematics}\\
{\it Massachusetts Institute of Technology},
{\it Cambridge, MA 02139} \\
\ \\
Gang ~He\\
{\it Department of Mechanical Engineering}\\
{\it Massachusetts Institute of Technology},
{\it Cambridge, MA 02139} }
\date{\today}

\begin{document}

\maketitle


\begin{abstract}
We present predictions for the statistical error due to finite
sampling in the presence of thermal fluctuations in molecular
simulation algorithms. Specifically, we establish how these errors
depend on Mach number, Knudsen number, number of particles, etc.
Expressions for the common hydrodynamic variables of interest such as
flow velocity, temperature, density, pressure, shear stress and heat
flux are derived using equilibrium statistical mechanics. Both
volume-averaged and surface-averaged quantities are
considered. Comparisons between theory and computations using direct
simulation Monte Carlo for dilute gases, and molecular dynamics for
dense fluids, show that the use of equilibrium theory provides
accurate results.
\end{abstract}

\small\normalsize
\section{Introduction}
Recently much attention has been focused on the simulation of
hydrodynamic problems at small scales using molecular simulation
methods such as Molecular Dynamics (MD) \cite{tilde,frenkel} or the direct
simulation Monte Carlo (DSMC) \cite{bird,dsmcintro}. Molecular
Dynamics is generally used to simulate liquids while DSMC is a
very efficient algorithm for simulating dilute gases. In molecular
simulation methods the connection to macroscopic observable
fields, such as velocity and temperature, is achieved through
averaging appropriate microscopic properties. The simulation results are therefore inherently statistical
and statistical errors due to finite sampling need to be fully
quantified.

Though confidence intervals may be estimated by measuring the
variance of these sampled quantities, this additional computation
can be burdensome and thus is often omitted. Furthermore, it would
be useful to estimate confidence intervals \textit{a priori} so
that one could predict the computational effort required to
achieve a desired level of accuracy. For example, it is well known
that obtaining accurate hydrodynamic fields (e.g., velocity
profiles) is computationally expensive in low Mach number flows so
it is useful to have an estimate of the computational effort
required to reach the desired level of accuracy.

In this paper we present expressions for the magnitude of
statistical errors due to thermal fluctuations in molecular
simulations for the typical observables of interest, such as
velocity, density, temperature, and pressure. We also derive
expressions for the shear stress and heat flux in the dilute gas
limit. Both volume averaging and
flux averaging is considered, even though the measurement of the
shear stress and heat flux through volume averaging is not exact
unless the contribution of the impulsive collisions is accounted
for. Although we make use of expressions from equilibrium
statistical mechanics, the non-equilibrium modifications to these
results are very small, even under extreme
conditions~\cite{nefluct}. This is verified by the good agreement
between our theoretical expressions and the corresponding
measurements in our simulations.

In addition to direct measurements of hydrodynamic fields,
this analysis will benefit algorithmic applications by providing a
framework were statistical fluctuations can be correctly accounted
for. One example of such application is the measurement of
temperature for temperature-dependent collision
rates~\cite{cuba}; another example is the measurement of velocity
and temperature for the purpose of imposing boundary
conditions~\cite{fang}. Additional examples include hybrid methods
\cite{nic,alex} where the coupling between continuum and molecular
fields requires both averaging of finite numbers of particles in a
sequence of molecular realizations as well as the generation of
molecular realizations, based on continuum fields, with relatively
small numbers of particles (e.g., buffer cells).

In section 2 the theoretical expressions for the statistical error
due to thermodynamic fluctuations are derived. These expressions
are verified by molecular simulations, as described in section 3.
The effect of correlations between samples in dilute gases is briefly discussed in section 4 and concluding remarks appear in section 5.

\section{Statistical error due to thermal fluctuations}
\label{theorysection}
\subsection{Volume-averaged quantities}
We first consider the fluid velocity. In a particle simulation,
the flow field is obtained by measuring the instantaneous center
of mass velocity, $\mathbf{u}$, for particles in a statistical
cell volume. The statistical mean value of the local fluid
velocity, $\langle \mathbf{u} \rangle_s$, is estimated over $M$
independent samples. For steady flows, these may be sequential
samples taken in time; for transient flows these may be samples
from an ensemble of realizations. The average fluid velocity,
$\langle \mathbf{u} \rangle$, is defined such that $\langle
\mathbf{u} \rangle_s \rightarrow \langle \mathbf{u} \rangle$ as $M
\rightarrow \infty$; for notational convenience we also write
$\langle \mathbf{u} \rangle = \mathbf{u}_0$.
Define $\delta u_x \equiv u_x - u_{x0}$ to be the instantaneous
fluctuation in the $x$-component of the fluid velocity; note that
all three components are equivalent. From equilibrium statistical
mechanics~\cite{landau},
\begin{equation}
\langle \delta u_x^2 \rangle = \frac{kT_0}{mN_0} =
\frac{a^2}{\gamma \mathrm{Ac}^2 N_0}
\label{deltau2Eqn}
\end{equation}
where $N_0$ is the average number of particles in the statistical
cell, $T_0$ is the average temperature, $m$ is the particle mass,
$k$ is Boltzmann's constant, $a$ is the sound speed, and $\gamma =
c_P/c_V$ is the ratio of the specific heats. The acoustic number
$\mathrm{Ac} = a/a^i$ is the ratio of the fluid's sound speed to
the sound speed of a ``reference'' ideal gas at the same
temperature
\begin{equation}
a^i = \sqrt{\gamma kT/m} \label{idealgas}
\end{equation}
Note that this reference ideal gas has a ratio of specific heats
($\gamma^i$) equal to the {\it original fluid} specific heat
ratio, that is $\gamma^i=\gamma$ as shown in equation
(\ref{idealgas}).

An alternative construction of (\ref{deltau2Eqn}) is obtained from the
equipartition theorem~\cite{landau}
\begin{equation}
{\textstyle \frac{3}{2}}kT_0={\textstyle \frac{1}{2}} m \langle
|\mathbf{c}-\mathbf{c}_0|^2 \rangle = {\textstyle \frac{1}{2}} m
\langle( (c_x-c_{x0})^2+(c_y-c_{y0})^2+(c_z-c_{z0})^2)\rangle
={\textstyle \frac{3}{2}} m \langle (c_x-c_{x0})^2 \rangle
\label{tequation}
\end{equation}
where $\mathbf{c}$ is the translational molecular velocity. For a
non-equilibrium system, this expression defines $T_0$ as the
average translational temperature. Note that $\mathbf{u}_0 =
\mathbf{c}_0$ and
\begin{equation}
\langle |\delta \mathbf{u}|^2 \rangle
= \frac{\langle | \mathbf{c}-\mathbf{c}_0|^2 \rangle}{N_0} = \frac{3kT_0}{mN_0}.
\end{equation}
The above expression also reminds us that the
expected error in estimating the magnitude of the fluid velocity
is $\sqrt{3}$ larger than in estimating a velocity component.

We may define a ``signal-to-noise'' ratio as the average fluid velocity
over its standard deviation; from the above,
\begin{equation}
\frac{|u_{x0}|}{\sqrt{\langle \delta u_x^2 \rangle}} = \mathrm{Ac}\,\mathrm{Ma}
\sqrt{\gamma N_0}
\end{equation}
where $\mathrm{Ma} = |u_{x0}|/a$ is the local Mach number based on the
velocity component of interest. This result shows that for fixed
Mach number, in a dilute gas simulation ($\mathrm{Ac}=1$), the
statistical error due to thermal fluctuations cannot be
ameliorated by reducing the temperature. However, when the Mach
number is small enough for compressibility effects to be
negligible, favorable relative statistical errors may be obtained
by performing simulations at an increased Mach number (to a level
where compressibility effects are still negligible).

The one-standard-deviation error bar for the sample estimate
$\langle u_x \rangle_s$ is $\sigma_u = \sqrt{\langle \delta u_x^2 \rangle}
/\sqrt{M}$ and the fractional error in the estimate of the fluid
velocity is
\begin{equation}
E_u = \frac{\sigma_u}{|u_{x0}|} = \frac{1}{\sqrt{MN_0}}\,\frac{1}
{\mathrm{Ac}\,\mathrm{Ma} \sqrt{\gamma}},
\label{velfluct}
\end{equation}
yielding
\begin{equation}
M = \frac{1}{\gamma \mathrm{Ac}^2 N_0 \mathrm{Ma}^2 E_u^2}.
\end{equation}
For example, with $N_0 = 100$ particles in a statistical cell, if
a one percent fractional error is desired in a $\mathrm{Ma}=1$
flow, about $M=100$ independent statistical samples are required
(assuming $\mathrm{Ac}\approx 1$). However, for a $\mathrm{Ma} =
10^{-2}$ flow, about $10^6$ independent samples are needed,
which quantifies the empirical observation that the resolution of the flow
velocity is computationally expensive for low Mach number flows.

Next we turn our attention to the density. From equilibrium statistical
mechanics, the fluctuation in the number of particles in a cell is
\begin{equation}
\langle \delta N^2 \rangle = - N^2 \frac{kT_0}{V^2}
\left(\frac{\partial V}{\partial P}\right)_T = \kappa_T N_0^2
\frac{kT_0}{V}
\end{equation}
where $V$ is the volume of the statistics cell and $\kappa_T \equiv
-V^{-1}(\partial V/\partial P)_T$ is the isothermal
compressibility. Note that for a dilute gas $\kappa_T = 1/P$ so
$\langle \delta N^2 \rangle = N$ and, in fact, $N$ is Poisson random
variable. The fractional error in the
estimate of the density is
\begin{equation}
E_\rho = \frac{\sigma_\rho}{\rho_0} = \frac{\sigma_N}{N_0} =
\frac{\sqrt{\langle \delta N^2 \rangle}}{N_0\sqrt{M}} =
\frac{\sqrt{\kappa_T k T_0}}{\sqrt{M V}} =
\frac{\sqrt{\kappa_T/\kappa_T^i}}{\sqrt{M N_0}}
\end{equation}
where $\kappa_T^i = V/N_0 kT_0$ is the isothermal compressibility
of the reference dilute gas ($\gamma^i=\gamma$) at the same
density and temperature. Since $a \propto 1/\sqrt{\kappa_T}$,
\begin{equation}
E_\rho = \frac{1}{\sqrt{M N_0}}\,\frac{1}{\mathrm{Ac}}
\label{densfluct}
\end{equation}
Note that for fixed $M$ and $N_0$, the error decreases as the
compressibility decreases (i.e., as the sound speed increases)
since the density fluctuations are smaller.

Let us now consider the measurement of temperature. First we should
remark that the measurement of instantaneous temperature is subtle,
even in a dilute gas. But given that temperature is measured correctly,
equilibrium statistical mechanics gives the variance in the temperature
fluctuations to be
\begin{equation}
\langle \delta T^2 \rangle = \frac{kT_0^2}{c_V N_0}
\end{equation}
where $c_V$ is the heat capacity per particle at constant volume.
The fractional error in the estimate of the temperature is
\begin{equation}
E_T = \frac{\sigma_T}{T_0}
= \frac{\sqrt{\langle \delta T^2 \rangle}}{T_0\sqrt{M}}
= \frac{1}{\sqrt{M N_0}}\,\sqrt{\frac{k}{c_V}}
\label{tempfluct}
\end{equation}
Because the fluctuations are smaller, the error in the temperature
is smaller when the heat capacity is large. Note that the
temperature associated with various degrees of freedom
(translational, vibrational, rotational) may be separately defined
and measured. For example, if we consider only the measurement of the
translational temperature, then the appropriate heat capacity is
that of an ideal gas with three degrees of freedom, i.e. $c_V =
\frac{3}{2}k$, corresponding to the three translational
components.

Finally, the variance in the pressure fluctuations is
\begin{equation}
\langle \delta P^2 \rangle = - kT_0 \left(\frac{\partial P}{\partial V}\right)_S
= \frac{\gamma k T_0}{V \kappa_T}
\end{equation}
so the fractional error in the estimate of the pressure is
\begin{equation}
E_P = \frac{\sigma_P}{P_0}
= \frac{\sqrt{\langle \delta P^2 \rangle}}{P_0\sqrt{M}}
= \frac{P_0^i}{P_0}\,\frac{\mathrm{Ac}\sqrt{\gamma}}{\sqrt{M N_0}}
\label{pressfluct}
\end{equation}
where $P_0^i = N_0 k T_0/V$ is the pressure of an ideal gas under
the same conditions. Note that the error in the pressure is
proportional to the acoustic number while the error in the
density, eqn.~(\ref{densfluct}), goes as $\mathrm{Ac}^{-1}$.

\subsection{Shear stress and heat flux for dilute gases}
\label{tauqSection}

The thermodynamic results in the previous section are general; in
this section we consider transport quantities and restrict our
analysis to dilute gases. In a dilute gas, the shear stress and
heat flux are defined as
\begin{equation}
\tau_{xy0} = \langle \tau_{xy}\rangle = \langle \rho c_x c_y \rangle
\end{equation}
and
\begin{equation}
q_{x0} = \langle q_{x}\rangle = \langle \frac{1}{2}\rho c_x c^2 \rangle
\end{equation}
respectively. This definition neglects the contribution of
impulsive collisions to the transport within a given volume due to
the negligible size of particles in the dilute limit. When,
however, simulation methods such as DSMC are used to simulate a
dilute gas that use a finite molecular size, a small error arises
when momentum and energy transport is calculated in a volume
averaged manner. In fact, this inconsistency has been addressed at
the equation of state level by Alexander et al.~\cite{cba}  who
pointed out that DSMC reproduces an ideal gas equation of state
while simulating transport of a hard-sphere gas with molecules of
finite size. Thus, under the assumption of a very dilute gas, the
fluctuation in the (equilibrium) shear stress and heat flux in a
volume $V$ containing $N_0$ particles can be calculated using the
Maxwell-Boltzmann distribution. In equilibrium,
the expected values of the above quantities are zero.

Using the definitions of shear stress and heat flux in terms of
moments of the velocity distribution, direct calculation of the
variance of the
$x$-$y$ component of the stress tensor based on a single-particle distribution function gives
\begin{eqnarray}
\langle \tau_{xy}^2\rangle&=&\langle (\rho c_x c_y)^2 \rangle \\
&=& \rho_0^2\; \langle (c_x c_y)^2 \rangle\\
&=& \frac{1}{4} \rho_0^2 \left(\frac{2 k T_0}{m}\right)^2 = P_0^2
\label{tauVarEqn1}
\end{eqnarray}

In obtaining the second equation we assumed $\langle c_x
\rangle=\langle c_y \rangle=\langle (\rho-\rho_0) c_x c_y
\rangle=0$. For the $x$ component of the heat flux vector, we find
\begin{eqnarray}
\langle q_x^2\rangle&=&\langle (\frac{1}{2}\rho c_x c^2)^2 \rangle \\
&=&\rho_0^2\;\langle (\frac{1}{2}c_x c^2)^2 \rangle \\
&=&\frac{35}{32} \rho_0^2 \left(\frac{2 k T_0}{m}\right)^3
=\frac{35}{8} c_\mathrm{m}^2 P_0^2 \label{qVarEqn1}
\end{eqnarray}
where $c_\mathrm{m} = \sqrt{2 k T_0/m}$ the most probable particle
speed  and we have assumed
$\langle c_x \rangle=\langle c_y \rangle=\langle (\rho-\rho_0) c^2
\rangle=0$. Note that in equilibrium for a cell containing $N_0$ particles, the variance of the mean is given by
$\langle \delta \tau_{xy}^2
\rangle = \langle \tau_{xy}^2 \rangle/N_0$ and $\langle \delta q_x^2
\rangle = \langle q_x^2 \rangle/N_0$ for the shear stress and heat flux respectively.

In order to derive expressions for the relative fluctuations we need
expressions for the magnitude of the fluxes. We are only able to
provide closed form expressions for the latter in the continuum regime
where
\begin{equation}
\tau_{xy}=\mu \left(\frac{\partial u_x}{\partial y}+\frac{\partial
u_y}{\partial x}\right)
\label{tauEqn}
\end{equation}
and
\begin{equation}
q_x=-\kappa \frac{\partial T}{\partial x}
\label{qEqn}
\end{equation}
where $\mu$ is the coefficient of viscosity and $\kappa$ is the
thermal conductivity. Above Knudsen numbers of $\mathrm{Kn} \approx
0.1$, it is known that these continuum expressions are only
approximate and better results are obtained using more complicated
formulations from kinetic theory (e.g., Burnett's formulation).  Here,
the Knudsen number is defined as $\mathrm{Kn}=\lambda/\ell$ and the
mean free path as
\begin{equation}
\lambda = \frac{8}{5\sqrt{ \pi}}  \frac{c_\mathrm{m} \mu}{P_0}
\label{meanfreepathdef}
\end{equation}
Note that this expression for the mean free path simplifies to the
hard sphere result when the viscosity is taken to be that of hard
spheres.

Using (\ref{tauEqn}) and (\ref{qEqn}) we find that in \emph{continuum
flows}, the relative fluctuations in the shear stress and heat flux
are given by
\begin{equation}
E_\tau=\frac{\sqrt{\langle \delta
\tau_{xy}^2\rangle}}{|\tau_{xy0}|\sqrt{M}} = \frac{16}{5\sqrt{2 \pi
\gamma}} \frac{1}{\mathrm{Kn}\,\mathrm{Ma}_*} \frac{1}{\sqrt{N_0
M}}\left|\frac{\partial u_x^*}{\partial y^*}+\frac{\partial
u_y^*}{\partial x^*}\right|^{-1} \label{stressfluct}
\end{equation}
and
\begin{equation}
E_q=\frac{\sqrt{\langle \delta q_x^2\rangle}}{|q_{x0}|\sqrt{M}}
=\frac{8\sqrt{35}}{5 \sqrt{2\pi}}
\frac{\mathrm{Pr}(\gamma-1)}{\gamma}\frac{T}{\Delta
T}\frac{1}{\mathrm{Kn}}\frac{1}{\sqrt{N_0 M}}\left|\frac{\partial
T^*}{\partial x^*}\right|^{-1} \label{heatfluct}
\end{equation}
respectively. Here, stars denote non-dimensional quantities:
$u^*=u_0/\tilde{u}$, $T^*=T/\Delta T$ and $x^*=x/\ell$, where
$\tilde{u}$, $\Delta T$, and $\ell$ are characteristic velocity,
temperature variation and length. The Mach number $\mathrm{Ma}_*$ is
defined with respect to the characteristic velocity $\tilde{u}$ rather
than the local velocity as in eq. (\ref{velfluct}).

If viscous heat generation is responsible for the temperature
differences characterized by $\Delta T$, then it is possible to
express equation (\ref{heatfluct}) in the following form
\begin{equation}
E_q=\frac{\sqrt{\langle \delta q_x^2\rangle}}{|q_{x0}|\sqrt{M}}=
\frac{8\sqrt{35}}{5\gamma
\sqrt{2\pi}}\frac{\mathrm{Br}}{\mathrm{Kn}
\mathrm{Ma}^2_*}\frac{1}{\sqrt{N_0 M}}\left|\frac{\partial
T^*}{\partial x^*}\right|^{-1} \label{stressfluct2}
\end{equation}
The Brinkman number
\begin{equation}
\mathrm{Br}=\frac{\mu \tilde{u}^2}{\kappa \Delta T},
\end{equation}
is the relevant non-dimensional group that compares temperature
differences due to viscous heat generation to the characteristic
temperature differences in the flow. (It follows that if viscous heat
generation is responsible for the temperature changes,
$\mathrm{Br}\approx 1$.)

It is very instructive to extend the above analysis to equation
(\ref{tempfluct}). If we define the relative error in temperature
with respect to the temperature changes rather than the absolute
temperature, we obtain
\begin{eqnarray}
E_{\Delta T} &=& \frac{\sigma_T}{\Delta T}
= \frac{\sqrt{\langle \delta T^2 \rangle}}{\Delta T \sqrt{M}}
= \frac{T_0}{\Delta T\sqrt{M N_0}}\,\sqrt{\frac{k}{c_V}}\label{tempfluct2}\\
 &=& \frac{1}{\mathrm{Pr}(\gamma-1)}\frac{\mathrm{Br}}{\mathrm{Ma}_*^2}\frac{1}{\sqrt{M N_0}}\,\sqrt{\frac{k}{c_V}}
\end{eqnarray}
where again, if viscous heat generation is the only source of heat,
$\mathrm{Br}\approx 1$. The above development shows that resolving the
temperature {\it differences} or heat flux due to viscous heat
generation is very computationally inefficient for low speed flow
since for a given expected error $E_{\Delta T}$ we find that the
number of samples $M\propto \mathrm{Ma}^{-4}_*$.

Comparison of equations (\ref{velfluct}) and (\ref{stressfluct})
and equations (\ref{tempfluct}) and (\ref{heatfluct})  reveals
that
\begin{equation}
E_\tau \sim \frac{E_u}{\mathrm{Kn}}
\end{equation}
and
\begin{equation}
E_q \sim \frac{E_{\Delta T}}{\mathrm{Kn}}
\end{equation}
since the non-dimensional gradients will be of order one. As the above
equations were derived for the continuum regime ($\mathrm{Kn}<0.1$),
it follows that the relative error in these moments is significantly
higher. This will also be shown to be the case in the next section
when the shear stress and heat flux are evaluated as fluxal (surface)
quantities. This has important consequences in hybrid methods
\cite{alex,nic}, as coupling in terms of state (Dirichlet) conditions
is subject to less variation than coupling in terms of flux
conditions.

\subsection{Fluxal quantities}

In this section we predict the relative errors in the fluxes of mass
and heat as well as the components of the stress tensor, when
calculated as fluxes across a reference surface.  Our analysis is
based on the assumption of an infinite, ideal gas in equilibrium.  In
this case, it is well known that the number of particles, $N_i$, in
each infinitessimal cell $i$ is an independent Poisson random variable
with mean and variance,
\begin{equation}
\langle N_i \rangle = \langle \delta N_i^2 \rangle = n \Delta x_i \Delta
y_i \Delta z_i ,
\end{equation}
where $n = \rho/m$ is the particle number density. 

\begin{figure}
\begin{center}
\includegraphics[width=4.5in]{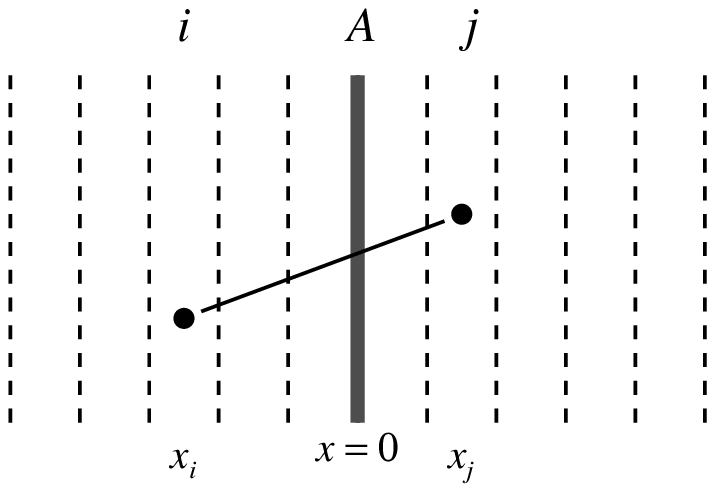}
\caption{ \label{geometryFig} Schematic particle motion crossing surface $A$.
}
\end{center}
\end{figure}

\begin{figure}
\begin{center}
\includegraphics[width=4.5in]{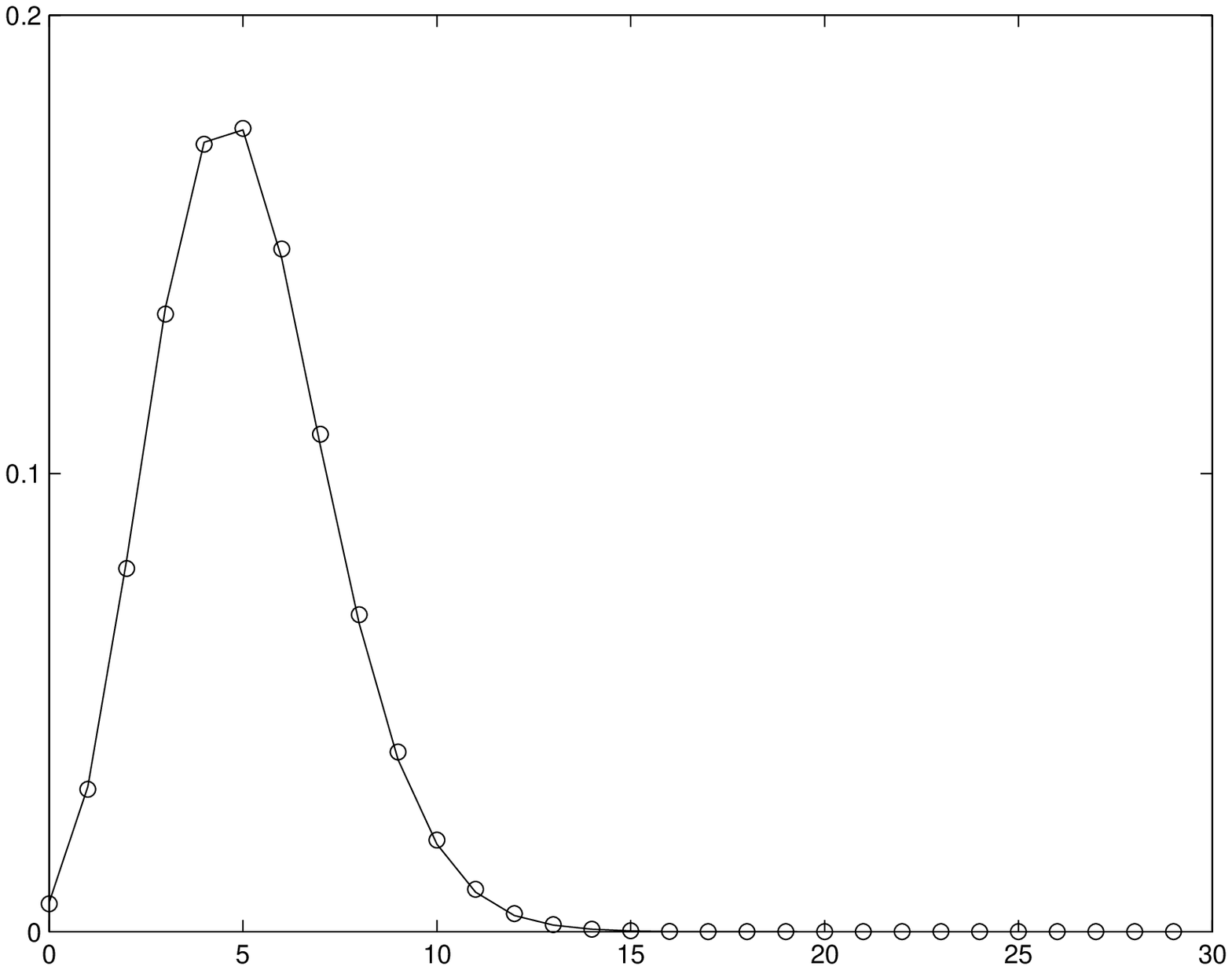}
\put(-40,-10){$N$} \put(-350,220){$P(N)$} \caption{\label{poisson}
Probability disrtibution of number of particles crossing a surface
in the gas. The circles denote equilibrium DSMC results and the
solid line shows a Poisson distribution with mean $\langle
N\rangle$ based on the equilibrium result $\langle J^+\rangle=n
c_\mathrm{m}/2\sqrt{\pi}$.}
\end{center}
\end{figure}

In the Appendix, it is shown that, more generally, all number
fluctuations are independent and Poisson distributed. In particular,
the number of particles, $N^+_{ij}$, leaving a cell $i$ at position
$(x_i,y_i,z_i)$, crossing a surface, and arriving in another cell $j$
at position $(x_j,y_j,z_j)$ after a time $\Delta t$ (see
Figure~\ref{geometryFig}) is also an independent Poisson random
variable. Its mean and variance are given by
\begin{equation}
\langle N_{ij}^+\rangle= \langle (\delta N_{ij}^+)^2 \rangle =
P_{ij} \langle N_i\rangle
\label{poissonstatement}
\end{equation}
where
\begin{equation}
P_{ij} = P(x_{ij},y_{ij},z_{ij})\Delta x_j \Delta y_j  \Delta z_j,
\end{equation}
is the probability that a particle makes the transition, which depends
only upon the relative displacements, $x_{ij} = x_j - x _i$, $y_{ij} =
y_j-y_i$, and $z_{ij} = z_j-z_i$. These properties hold exactly for an
infinite ideal gas in equilibrium, but they are also excellent
approximations for most finite, dilute gases, even far from
equilibrium, e.g. as shown by the DSMC result in Figure~\ref{poisson}.

Without loss of generality we place a large planar surface of area $A$ at
$x=0$. The flux in the positive direction, $J_h^+$, of any
velocity-dependent quantity, $h(c_x,c_y,c_z)$, can be expressed as a sum of
independent, random contributions from different cells $i$ and $j$ on
opposite sides of the surface,
\begin{equation}
J_h^+ =  \frac{1}{A \Delta t} \sum_{x_i<0} \sum_{x_j>0} \sum_{y_i}
\sum_{y_j} \sum_{z_i} \sum_{z_j} \bar{h}(x_{ij},y_{ij},z_{ij})  N_{ij}^+
\label{Jh}
\end{equation}
where $\bar{h}(x_{ij},y_{ij},z_{ij}) = h(c_x,c_y,c_z)$ is the same
quantity expressed in terms of relative displacements during the time
interval, $\Delta t$. We begin by calculating the mean flux. Taking
expectations in Eq.~(\ref{Jh}), we obtain
\begin{eqnarray}
\langle J^+_h \rangle &=&  \frac{1}{A \Delta t} \sum_{x_i<0} \sum_{x_j>0}
\sum_{y_i} \sum_{y_j} \sum_{z_i} \sum_{z_j} \bar{h}(x_{ij},y_{ij},z_{ij})
\langle N_i \rangle P_{ij} \label{Jhmean} \\ & \sim & \frac{1}{\Delta t}
\sum_{x_i <0} \sum_{x_j>0} \sum_{y_j} \sum_{z_j} \bar{h}(x_{ij},y_j,z_j)  n
\Delta x_i P(x_{ij},y_j,z_j) \Delta x_j \Delta y_j \Delta z_j,
\label{Jhmean2}
\end{eqnarray}
where without loss of generality we have taken $y_i=z_i=0$ and also separately performed the sums parallel to the
surface,
\[
\sum_{y_i} \sum_{z_i} \Delta y_i \Delta z_i = A,
\]
by invoking translational invariance and neglecting any edge effects,
since $\langle c_y \rangle \Delta t, \langle c_z \rangle \Delta t =
o(\sqrt{A})$ in the limit $\Delta t \rightarrow 0$ (taken
below). 

Passing to the continuum limit in Eq.~(\ref{Jhmean2}) and
letting $A \rightarrow \infty$, we arrive at an integral expression
for the mean flux through an infinite flat surface,
\begin{eqnarray}
\nonumber\langle J^+_h \rangle &=& \frac{n}{\Delta t} \;
\int^0_{-\infty} dx^\prime \; \int^{\infty}_0 dx
\;\int^{\infty}_{-\infty}dy \;\int^{\infty}_{-\infty}dz\;
\bar{h}(x-x^\prime,y,z) P(x-x^\prime,y,z) \\ &=& \frac{n}{\Delta t} \;
\int^{\infty}_0 dx \; \int_x^\infty ds \;\int^{\infty}_{-\infty}dy
\;\int^{\infty}_{-\infty}dz\; \bar{h}(s,y,z) P(s,y,z).
\end{eqnarray}
Since the integrand does not depend on $x$, switching the order of
integration,
\begin{eqnarray}
\nonumber\langle J^+_h \rangle &=& \frac{n}{\Delta t} \; \int_0^\infty ds
\;\int^{\infty}_{-\infty}dy \;\int^{\infty}_{-\infty}dz\; \bar{h}(s,y,z)
P(s,y,z) \int_0^s dx \\ &=& n \; \int_0^\infty ds
\;\int^{\infty}_{-\infty}dy \;\int^{\infty}_{-\infty}dz\; \bar{h}(s,y,z)
P(s,y,z) \frac{s}{\Delta t}
\end{eqnarray}
produces a simple formula for the mean flux as a conditional average
over the velocity distribution (with $c_x > 0$),
\begin{equation}
\langle J^+_h \rangle  = n \; \langle c_x\; h(c_x,c_y,c_z)\rangle^+
\label{Jhformula}
\end{equation}
in the limit $\Delta t \rightarrow 0$. For $h(c_x,c_y,c_z)=m,m c_x,m
c_y, m c_z, \frac{1}{2}mc^2$ in the absence of a mean flow, we obtain,
\begin{eqnarray}
\langle J_m^+ \rangle &=& n\langle c_x\;m\rangle^+ = \frac{\rho c_\mathrm{m}}{2 \sqrt{\pi}}\\
\langle J_{xx}^+ \rangle &=&  n\langle c_x\;m c_x \rangle^+ = \frac{\rho k T_0}{2 m} = \frac{P_0}{2}\\
\langle J_{xy}^+ \rangle &=&  n\langle c_x\;m c_y \rangle^+ = 0\qquad\qquad
\langle J_{xz}^+ \rangle =  n\langle c_x\;m c_z \rangle^+ = 0\\
\langle J_e^+ \rangle &=&  n\langle c_x\; {\textstyle
\frac{1}{2}}m c^2 \rangle^+ = \frac{\rho c_\mathrm{m}^3}{2
\sqrt{\pi}}
\end{eqnarray}
which are the well-known results for the expected one-sided fluxes
of mass, momentum, and energy, respectively.

\begin{figure}
\begin{center}
\includegraphics[width=4.1in]{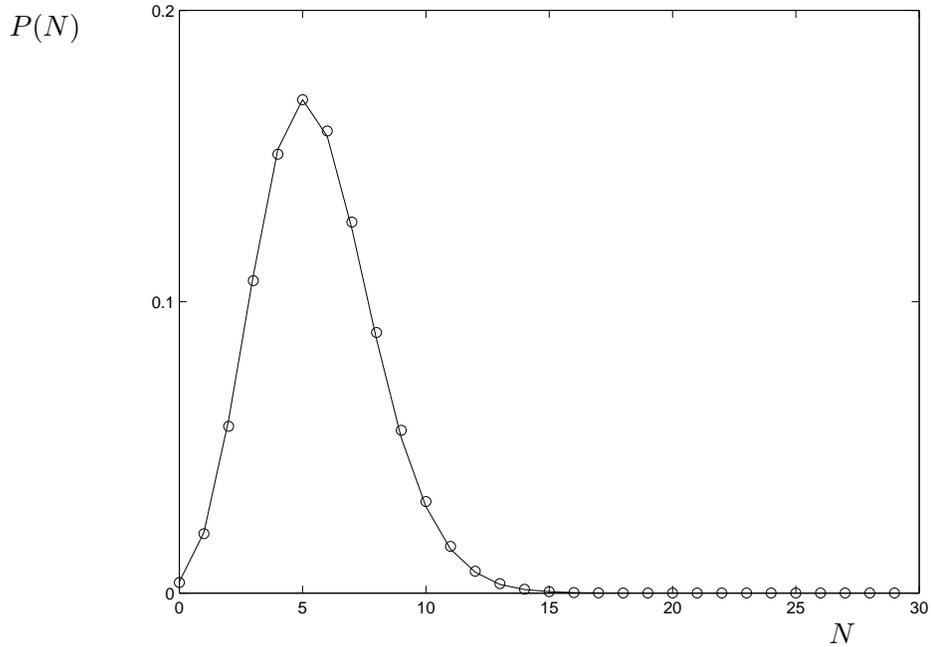}
\put(-40,-10){$N$} \put(-350,220){$P(N)$}
\caption{\label{poissonmean} Probability disrtibution of number of
particles crossing a surface in the gas in the presence of mean
flow normal to the surface of interest. The circles denote
equilibrium DSMC results and the solid line shows a Poisson
distribution with mean $\langle N\rangle$ based on the particle
flux in the presence of a mean flow.}
\end{center}
\end{figure}

Using the same formalism as above, we now derive a general formula for
the variance of the flux, $\langle (\delta J^+_h)^2 \rangle$. For
simplicity, we assume that the mean flow is zero normal to the
reference surface, but our results below are easily extended to the
case of a non-zero mean flow through the surface since
Eq.~(\ref{poissonstatement}) still holds in that case (see Figure
\ref{poissonmean}). Taking the variance of Eq.~(\ref{Jh}) and using
the independence of $\{ N_{ij}^+ \}$, we obtain
\begin{eqnarray}
\langle (\delta J^+_h)^2 \rangle &=&\frac{1}{A^2\Delta t^2}
 \sum_{x_i<0} \sum_{x_j>0} \sum_{y_i}
\sum_{y_j} \sum_{z_i} \sum_{z_j} \bar{h}(x_{ij},y_{ij},z_{ij})^2 \langle (\delta
N^+_{ij})^2 \rangle   \nonumber \\ &=&\frac{1}{A^2\Delta t^2}
 \sum_{x_i<0} \sum_{x_j>0} \sum_{y_i}
\sum_{y_j} \sum_{z_i} \sum_{z_j} \bar{h}(x_{ij},y_{ij},z_{ij})^2 \langle N_i
\rangle P_{ij}
\end{eqnarray}
Following the same steps above leading from Eq.~(\ref{Jhmean}) to
Eq.~(\ref{Jhformula}) we arrive at the simple formula,
\begin{equation}
\langle (\delta J^+_h)^2 \rangle  =\frac{n}{A\Delta t}\;\langle c_x\;
h(c_x,c_y,c_z)^2\rangle^+
\end{equation}
in the continuum limit as $\Delta t \rightarrow 0$. The variance of the
total flux is
\begin{equation}
\langle \delta J_h^2 \rangle = \langle (\delta J_h^+ + \delta J_h^-)^2
\rangle = \langle (\delta J_h^+)^2 \rangle + \langle (\delta J_h^-)^2
\rangle = 2 \langle (\delta J^+_h)^2 \rangle
\end{equation}
since the one-sided fluxes through the surface, $J^+_h$ and $J^-_h$, are
independent and identically distributed (with opposite sign).

Using this general result, we can evaluate the standard deviations of the
fluxes above,
\begin{eqnarray}
\sqrt{\langle \delta J_m^2 \rangle} &=&
\sqrt{\frac{\rho c_\mathrm{m}}{\sqrt{\pi}}}\sqrt{\frac{m}{A\Delta t}}\\[5pt]
\sqrt{\langle \delta J_{xy}^2 \rangle} &=&
\sqrt{\frac{\rho c_\mathrm{m}^3}{2\sqrt{\pi}}}\sqrt{\frac{m}{A\Delta t}}\\[5pt]
\sqrt{\langle \delta J_{e}^2 \rangle} &=& \sqrt{\frac{3\rho
c_\mathrm{m}^5}{2\sqrt{\pi}}}\sqrt{\frac{m}{A\Delta t}}
\end{eqnarray}
These formulae may be simplified by noting that
\begin{equation}
\langle J^+_m \rangle =\frac{mN^+}{A\Delta t}=\frac{\rho
c_\mathrm{m}}{2\sqrt{\pi}}
\end{equation}
where $N^+$ is the mean total number of particles crossing the reference
surface in one direction in time $\Delta t,$, which yields
\begin{eqnarray}
\sqrt{\langle \delta J_m^2 \rangle} &=&
\frac{\rho c_\mathrm{m}}{\sqrt{2\pi}}\frac{1}{\sqrt{N^+}}\\[5pt]
\sqrt{\langle \delta J_{xy}^2 \rangle} &=&
\frac{\rho c^2_\mathrm{m}}{2\sqrt{\pi}}\frac{1}{\sqrt{N^+}} \label{tauVarEqn2}\\[5pt]
\sqrt{\langle \delta J_{e}^2 \rangle}&=&\sqrt{\frac{3}{\pi}}\frac{\rho c^3_\mathrm{m}}{2} \frac{1}{\sqrt{N^+}}\label{qVarEqn2}
\end{eqnarray}
We may relate these to the results from the previous section by
identifying $\langle \delta (\tau_{xy}^f)^2\rangle = \langle \delta J_{xy}^2 \rangle$ and $\langle \delta (q_{x}^f)^2\rangle = \langle \delta J_{e}^2 \rangle$, and additionally include the effect of $M$ (independent) samples in time. The superscript $f$ denotes fluxal measurement. By noting that transport fluxes are defined with respect to the rest frame of the fluid, it can be easily verified that the above relations hold in the case where a mean flow in directions parallel to the measuring surface exists, under the assumption of a local equilibrium distribution.

We can derive expressions for the relative expected error in the
continuum regime in which models exist for the shear stress and
heat flux.  In this regime we find
\begin{eqnarray}
E_{\tau}^f=\frac{\sqrt{\delta (\tau_{xy}^f)^2}}{|\tau_{xy0}^f|\sqrt{M}}
&=&\frac{16}{5\pi \sqrt{2\gamma}}
\frac{1}{\mathrm{Kn}\;\mathrm{Ma}_*}\frac{1}{\sqrt{M N^+}}
\left|{\frac{\partial u^*}{\partial y^*}}+{\frac{\partial
v^*}{\partial x^*}}\right|^{-1} \label{stressfluxfluct}
\end{eqnarray}
and
\begin{eqnarray}
E_q^f=\frac{\sqrt{\delta (q_x^f)^2}}{|q_{x0}^f| \sqrt{M}}
&=&\frac{16\sqrt{3}}{5\pi}\frac{\mathrm{Pr}(\gamma-1)}{\gamma}\frac{T}{\Delta T}\frac{1}{\mathrm{Kn}}\frac{1}{\sqrt{M N^+}} \left|
 \frac{\partial T}{\partial x^*}\right|^{-1}\\[5pt]
&=&\frac{16\sqrt{3}}{5\gamma \pi}\frac{\mathrm{Br}}{\mathrm{Kn}\;\mathrm{Ma}^2_*}
\frac{1}{\sqrt{M N^+}}
\left|\frac{\partial T}{\partial x^*}\right|^{-1}\label{heatfluxfluct}
\end{eqnarray}
Comparing (\ref{stressfluxfluct}) with the corresponding expressions
for volume-averaged stress tensor, (\ref{stressfluct2}), one finds
that, aside from the numerical coefficients, the expressions differ
only in the number of particles used, either $N^+$ or $N_0$; one finds
a similar result for the heat flux.  
\subsection{Connection to Fluctuating Hydrodynamics}

Fluctuating hydrodynamics, as developed by Landau, approximates
the stress tensor and heat flux as white noises, with variances
fixed by matching equilibrium fluctuations. In this section we identify
the connection between Landau's theory and the variances of fluxes
obtained in section~\ref{tauqSection}.

Landau introduced fluctuations into the hydrodynamic equations by
adding white noise terms to the stress tensor and heat
flux~\cite{landau2} (in the spirit of Langevin's theory of Brownian
motion~\cite{langevin}). The amplitudes of these noises are fixed by
evaluating the resulting variances of velocity and temperature and
matching with the results from equilibrium statistical
mechanics~\cite{landau1}. For example, in Landau's formulation the
total heat flux in the $x$-direction is $q_x^\mathrm{L} = -\kappa
\partial T/\partial x + g_x$ where the first term on the r.h.s. is the
deterministic part of the flux and the second is the white noise
term. The latter has mean zero and time correlation given by the
following expression
\begin{equation}
\langle g_x(t) g_x(t') \rangle = \frac{ 2 k \kappa T^2 }{V} \delta (t - t')
\label{LandauQ}
\end{equation}
Note that at a steady state the deterministic part is constant so
$\langle \delta q_x^\mathrm{L}(t) \delta q_x^\mathrm{L}(t')
\rangle = \langle g_x(t) g_x(t') \rangle$.

On the other hand, recall from eqn.~(\ref{qVarEqn1}),
\begin{equation}
\langle \delta q_x^2 \rangle = \frac{35}{32} \frac{\rho^2
c_\mathrm{m}^6}{N_0} \label{KineticQ}
\end{equation}
The question naturally arises: how does one reconcile
(\ref{LandauQ}) and (\ref{KineticQ})? Note that the fluctuating
hydrodynamics expression contains the thermal conductivity, which
depends on the particle interaction (e.g., for hard-spheres
$\kappa$ depends on the particle diameter) while the kinetic
theory expression is independent of this interaction.

The key lies in identifying the $\delta$-function with a decay
time, $t_d$, that is vanishingly small at hydrodynamic scales.
Specifically, we may write
\begin{equation}
\delta ( t - t' ) \rightarrow \left\{
\begin{array}{cc}
t_d^{-1} & | t - t' | < t_d \\ 0 & \mathrm{otherwise}
\end{array} \right.
\end{equation}
so
\begin{equation}
\langle g^2_x(t) \rangle = \frac{ 2 k \kappa T^2 }{t_d V} =
\frac{\kappa V}{2 k N_0 c_\mathrm{m}^2 t_d} \frac{\rho^2
c_\mathrm{m}^6}{N_0}
\end{equation}
Comparing with the above gives
\begin{equation}
t_d = \frac{16}{35} \frac{\kappa V}{k N_0 c_\mathrm{m}^2}
\end{equation}
For a hard sphere gas, $\kappa = \frac{5}{2} c_V \mu$ so using
(\ref{meanfreepathdef}) we may write this as,
\begin{equation}
t_d = \frac{15\sqrt{\pi}}{28} \frac{\lambda}{c_\mathrm{m}}
\end{equation}
For other particle interactions the coefficients will be slightly
different but in general $t_d \approx \lambda/c_\mathrm{m}$, thus it
is approximately equal to the molecular collision time. In conclusion,
the two formulations are compatible once the white noise approximation
in fluctuating hydrodynamics is justified by the fact that the
hydrodynamic time scale is much longer than the kinetic (i.e.,
collisional) time scale. Landau's construction provides a useful
hydrodynamic approximation for $g_x$ but (\ref{KineticQ}) is the
actual variance of the heat flux.

\section{Simulations}
\subsection{Dilute Gases}
We performed DSMC simulations to verify the validity of the
expressions derived above. Standard DSMC techniques
\cite{bird,dsmcintro} were used to simulate flow of gaseous argon
(molecular mass $m=6.63\times 10^{-26}$~kg, hard sphere diameter
$\sigma=3.66\times 10^{-10}$~m) in a two-dimensional channel
(length $L$ and height $H$). The simulation was periodic in the
$x$ direction (along the channel axis). The two walls at $y=-H/2$
and $y=H/2$ were fully accommodating and flat. The simulation was
also periodic in the third (homogeneous) direction.

The average gas density was $\rho_0 = 1.78
\mathrm{kg}/\mathrm{m}^3$ and in all calculations over 40
particles per cell were used. The cell size was $\Delta
x=\lambda_0/3$ where $\lambda_0$ is the reference mean free path.
The time step was $\Delta t =\lambda_0/(7.5 c_\mathrm{m})$. For a
discussion of the errors resulting from finite cell sizes and time
steps see \cite{frankcell,NHtime,AGWWtime}. The fractional error
in the simulations is obtained from the standard deviation of cell
values in the $x$ and $z$ directions. To ensure that the samples
were independent, samples were taken only once every 250 time
steps. To ensure that the system was in its steady state the
simulation was run for $10^6$ time steps before sampling was
started.

\begin{figure}
\begin{center}
\includegraphics[width=4.1in]{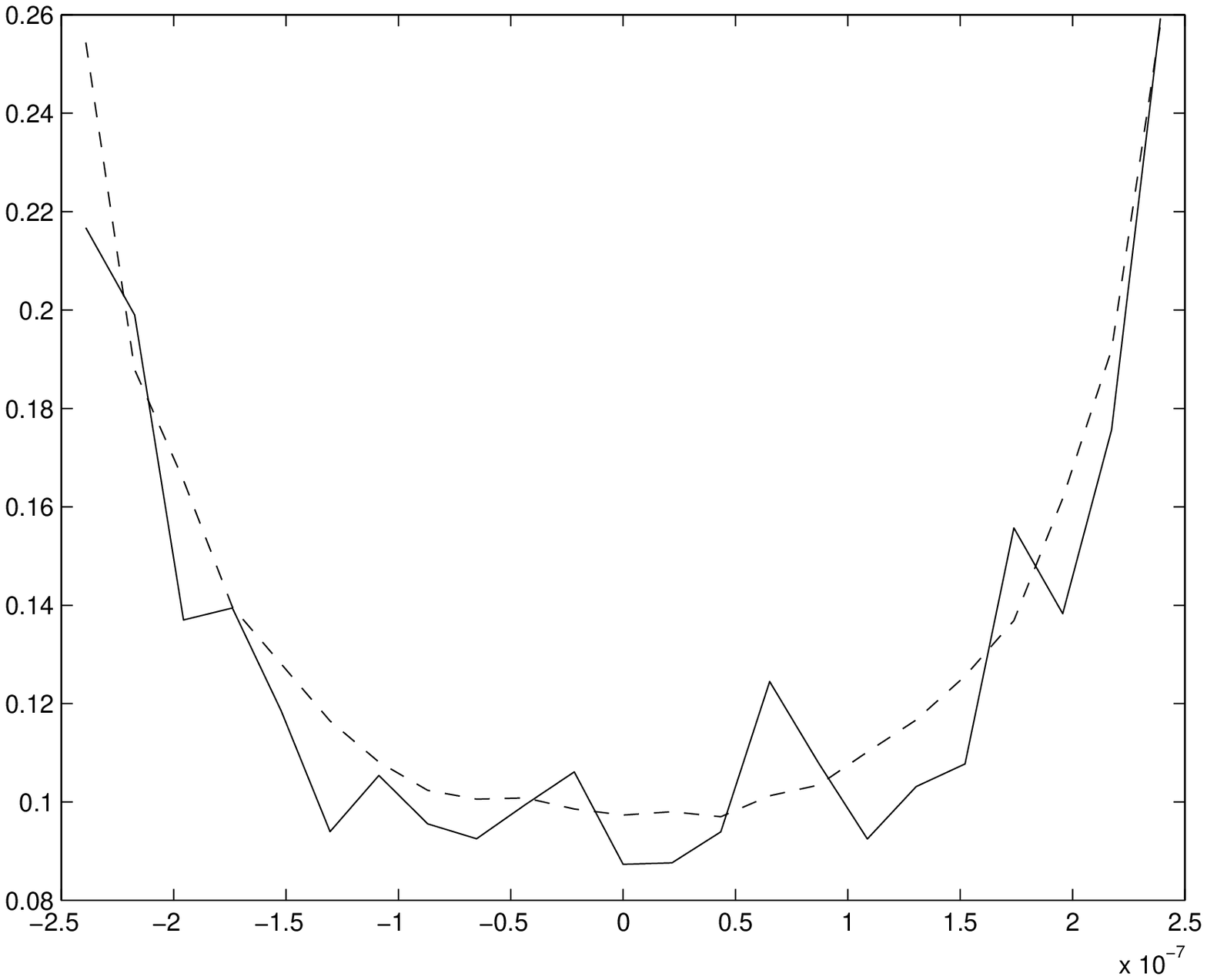}
\put(-40,-10){$y$} \put(-350,220){$E_u$} \caption{ \label{velocityfig}
Fractional error in velocity for Poiseuille flow in a channel as a
function of the transverse channel coordinate, $y$. The dashed
line denotes equation (\ref{velfluct}) and the solid line denotes
DSMC simulation results.}
\end{center}
\end{figure}

\begin{figure}
\begin{center}
\includegraphics[width=4.1in]{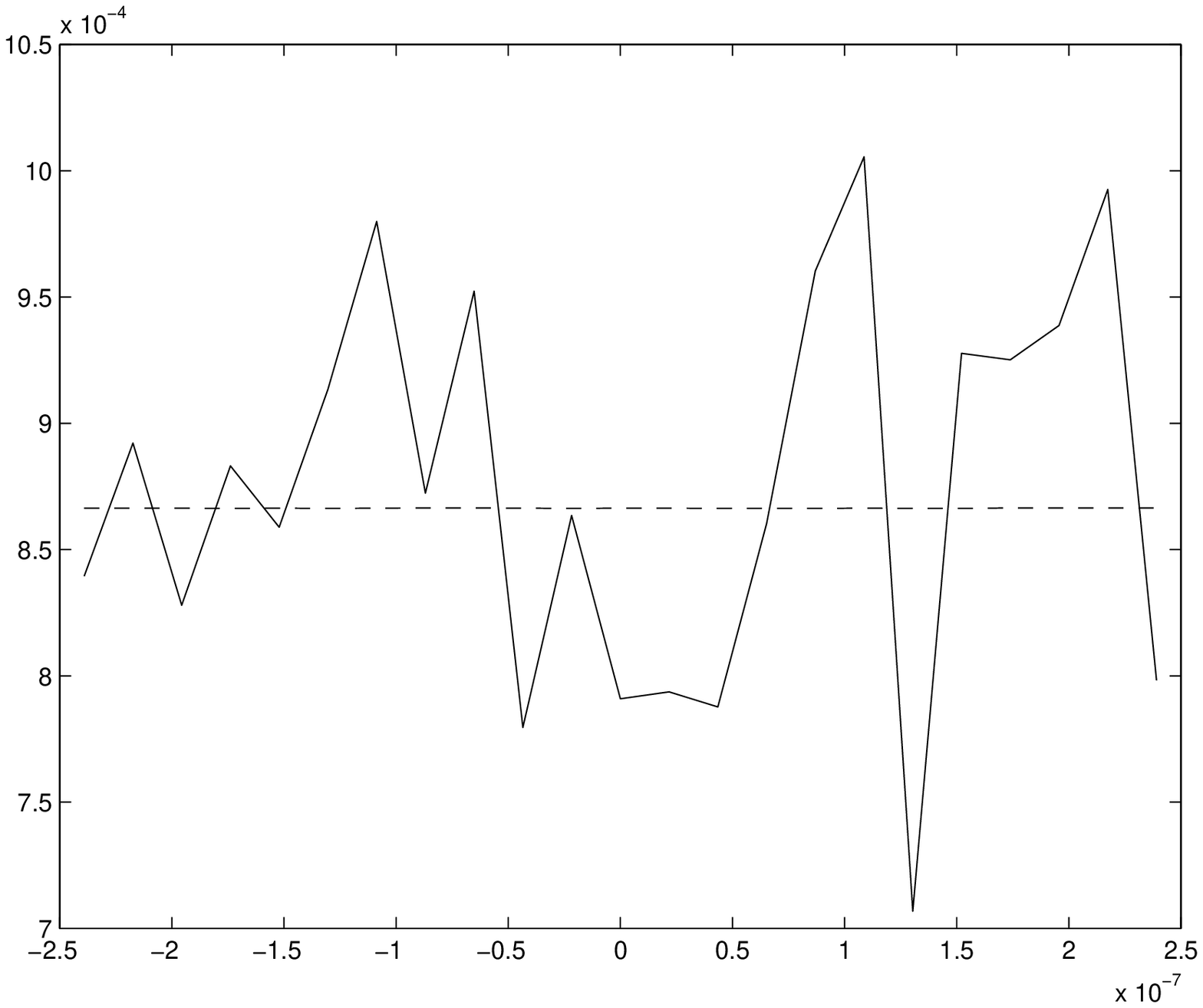}
\put(-40,-10){$y$} \put(-350,220){$E_{\rho}$} \caption{
\label{densityfig} Fractional error in density for Poiseuille flow in a
channel as a function of the transverse channel coordinate, $y$.
The dashed line denotes equation (\ref{densfluct}) and the solid
line indicates DSMC simulation results.}
\end{center}
\end{figure}

\begin{figure}
\begin{center}
\includegraphics[width=4.1in]{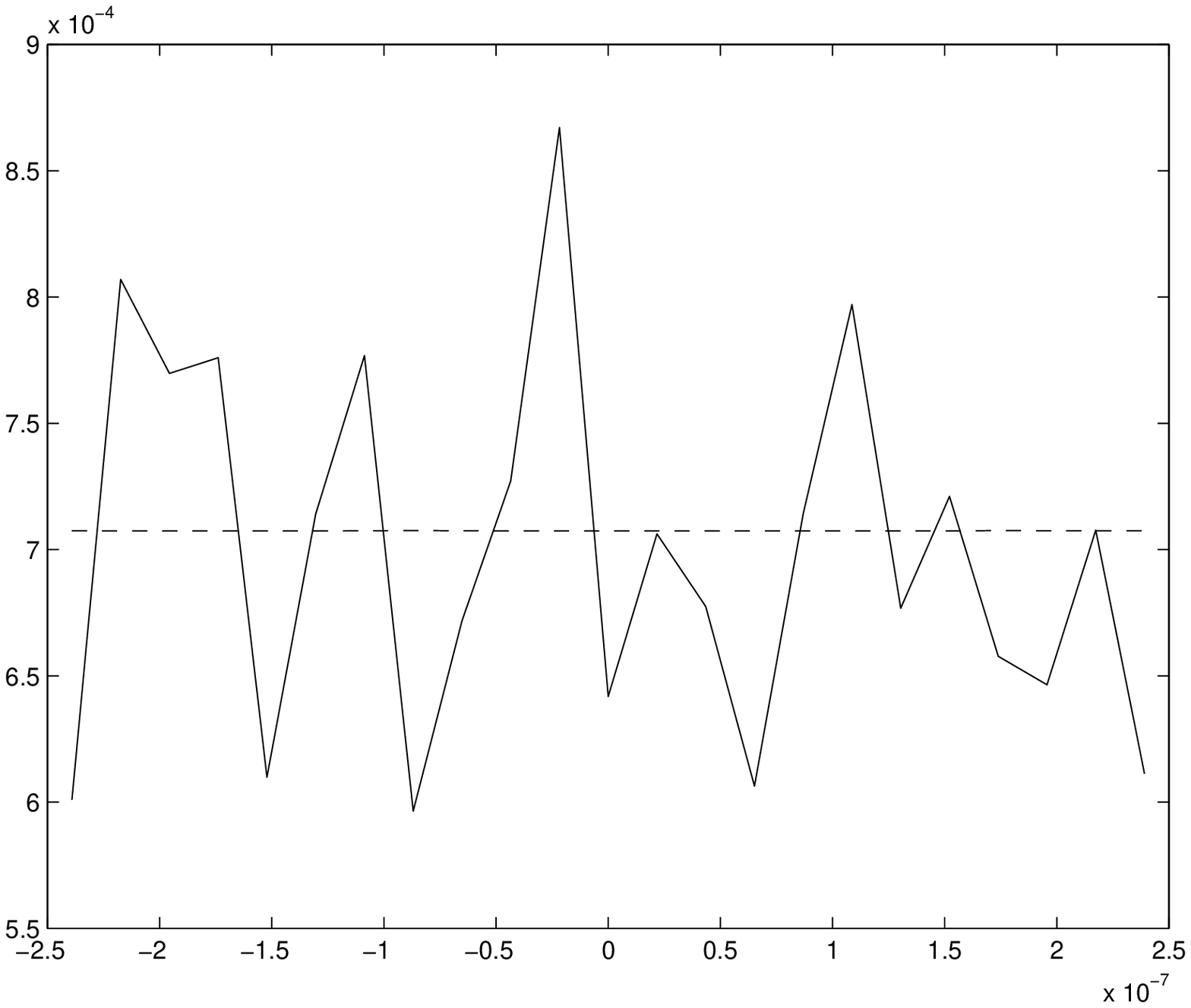}
\put(-40,-10){$y$} \put(-350,220){$E_T$} \caption{ \label{temperaturefig}
Fractional error in temperature for Poiseuille flow in a channel
as a function of the transverse channel coordinate, $y$. The
dashed line denotes equation (\ref{tempfluct}) and the solid line
indicates DSMC simulation results.}
\end{center}
\end{figure}

A constant acceleration was applied to the particles to produce
Poiseuille flow in the $x$ direction with maximum velocity at the
centerline $u_0^\mathrm{max} \approx 2 $~m/s. Figures
\ref{velocityfig}, \ref{densityfig}, and \ref{temperaturefig} show
good agreement between the theoretical expressions from section~2
and simulation measurements for the fractional error in velocity,
density and temperature, respectively. The fractional error in the
velocity measurement is minimum at the centerline since the
Poiseuille velocity profile is parabolic and maximum at the
centerline, (see Fig. \ref{velocityfig}). The density and
temperature were nearly constant across the system so the
fractional errors in these quantities are also nearly constant.

\begin{figure}
\begin{center}
\includegraphics[width=4.1in]{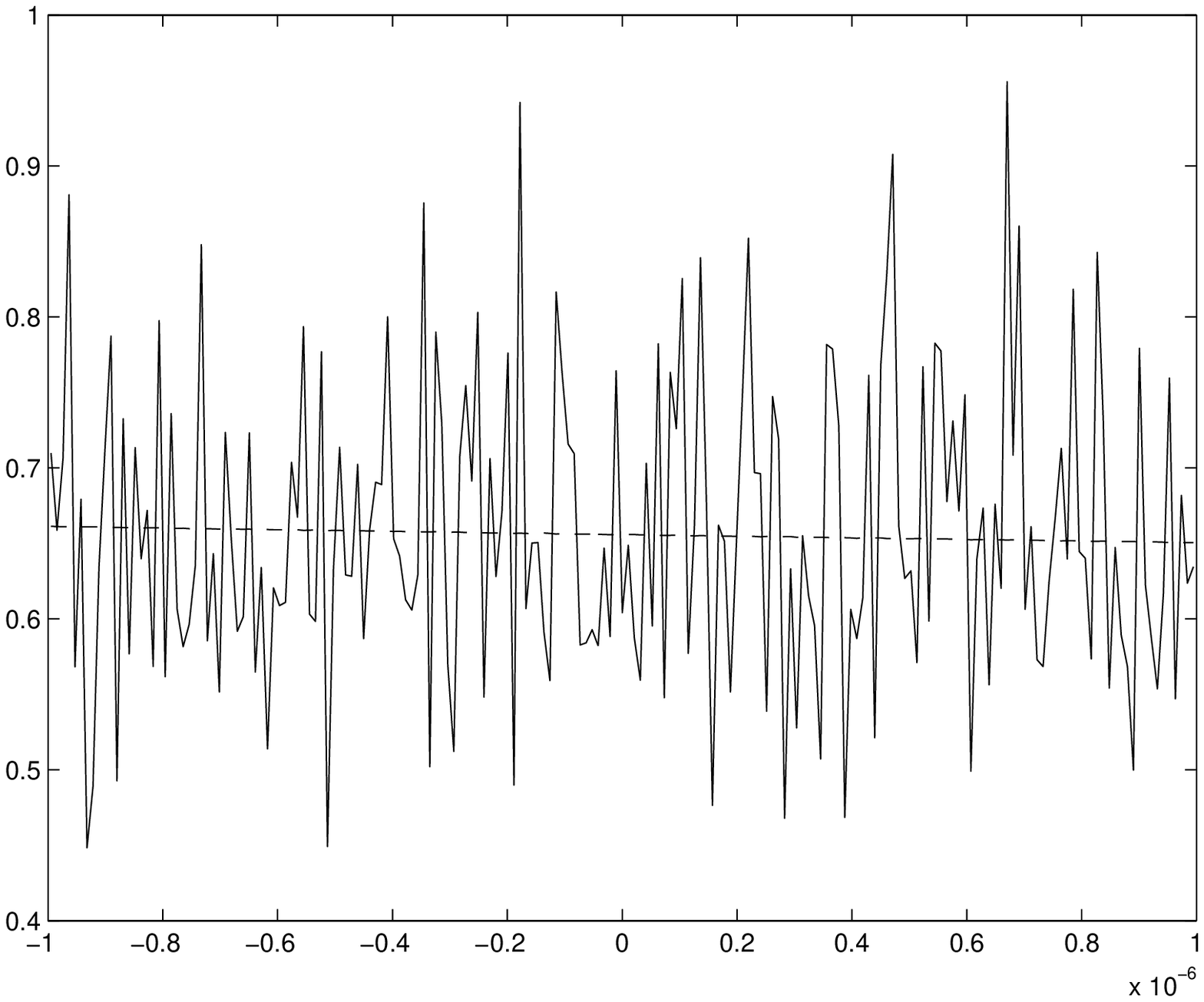}
\put(-40,-10){$y$} \put(-350,220){$E_{\tau}$} \caption{ \label{volumestress}Fractional error in
the shear stress $\tau_{xy}$ for Couette flow in a channel
as a function of the transverse channel coordinate, $y$. The
dashed line denotes equation (\ref{stressfluct}) and the solid line
indicates DSMC simulation results.}
\end{center}
\end{figure}

\begin{figure}
\begin{center}
\includegraphics[width=4.1in]{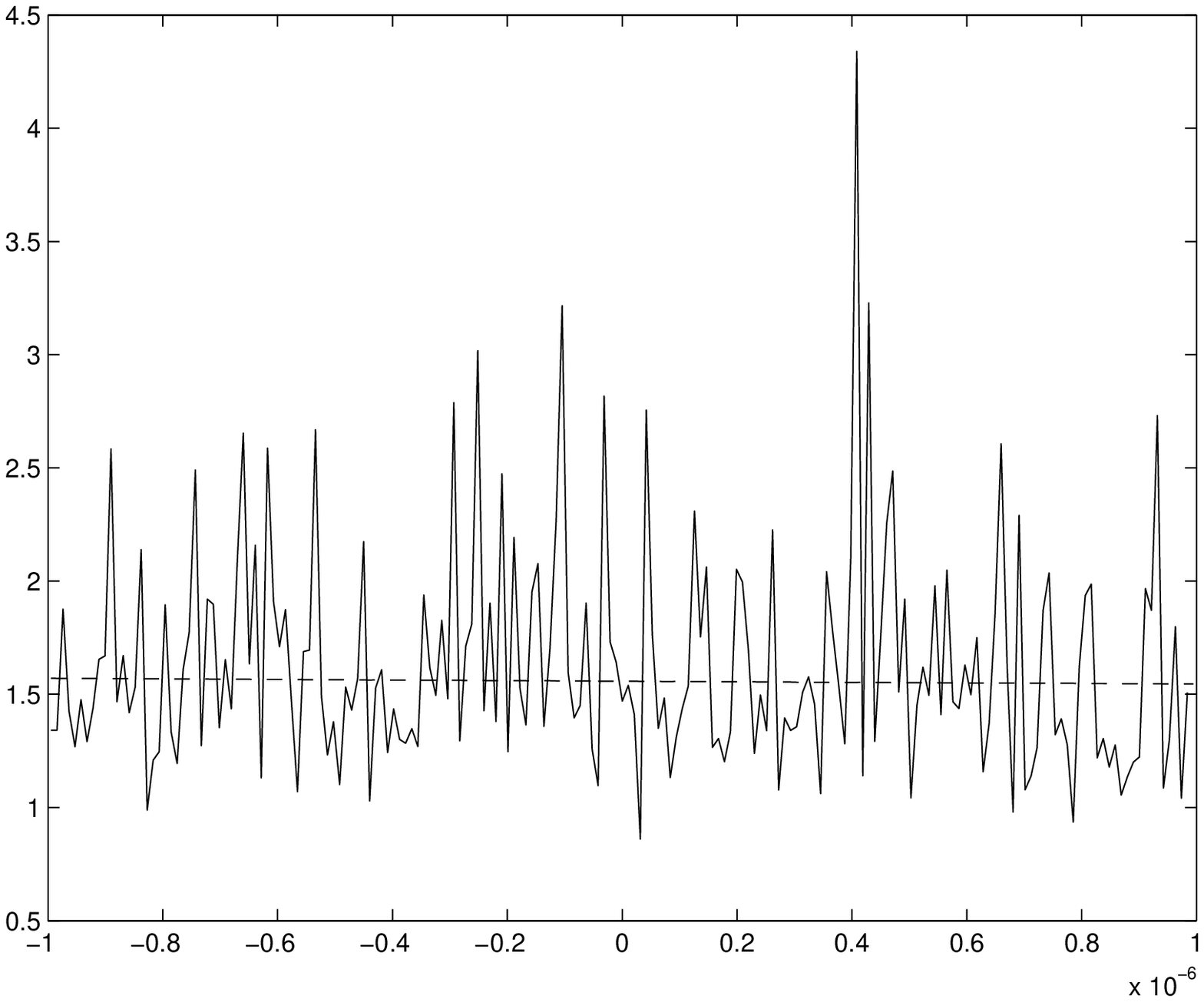}
\put(-40,-10){$y$} \put(-350,220){$E_{\tau}^f$} \caption{ \label{fluxstress}Fractional error in
the shear stress  for Couette flow in a channel
as a function of the transverse channel coordinate, $y$. The
dashed line denotes equation (\ref{stressfluxfluct}) and the solid line
indicates DSMC simulation results.}
\end{center}
\end{figure}

\begin{figure}
\begin{center}
\includegraphics[width=4.1in]{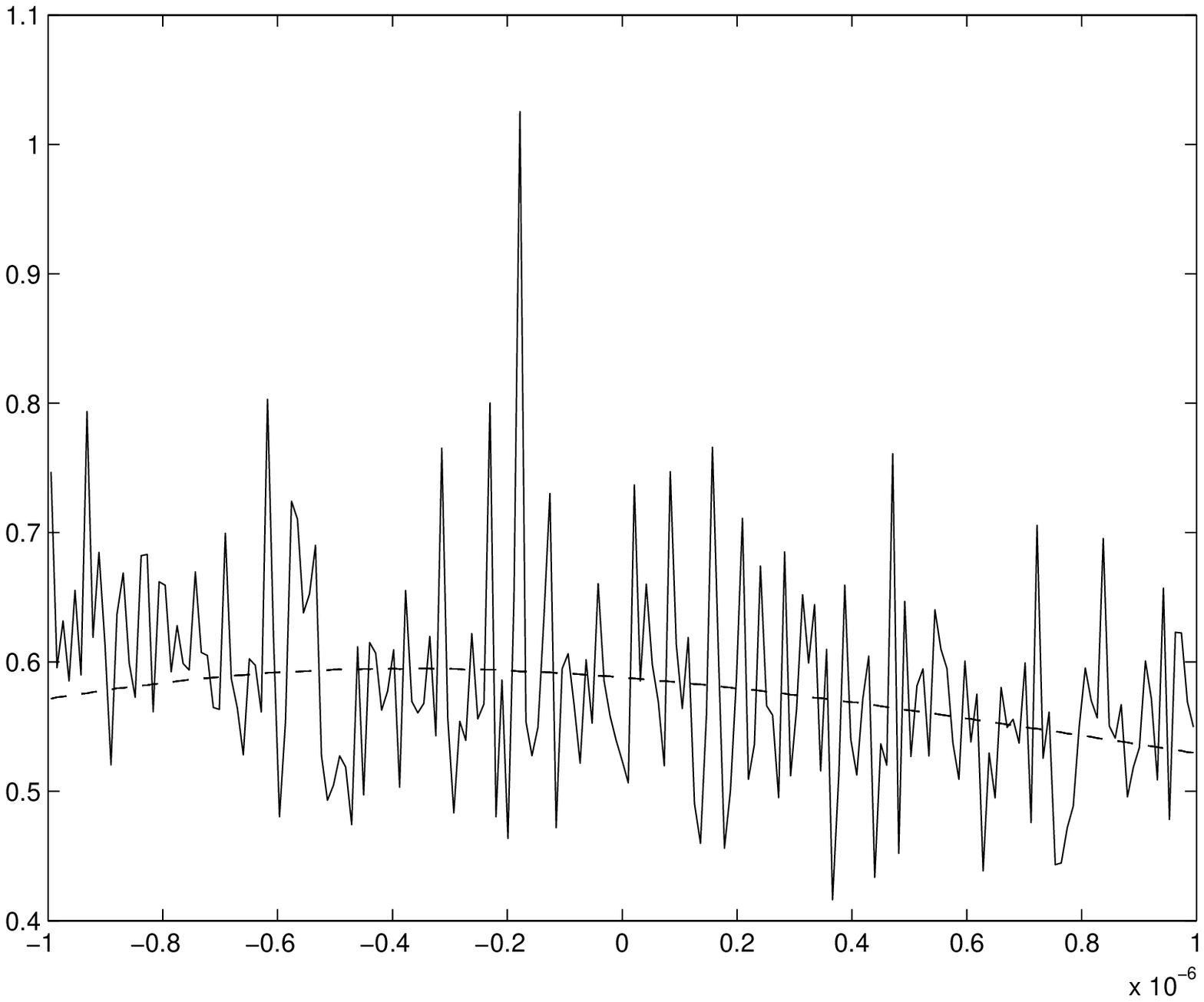}
\put(-40,-10){$y$} \put(-350,220){$E_q$} \caption{ \label{volumeflux}Fractional error in
the heat flux for ``Tempearture Couette'' in a channel
as a function of the transverse channel coordinate, $y$. The
dashed line denotes equation (\ref{heatfluct}) and the solid line
indicates DSMC simulation results.}
\end{center}
\end{figure}

\begin{figure}
\begin{center}
\includegraphics[width=4.1in]{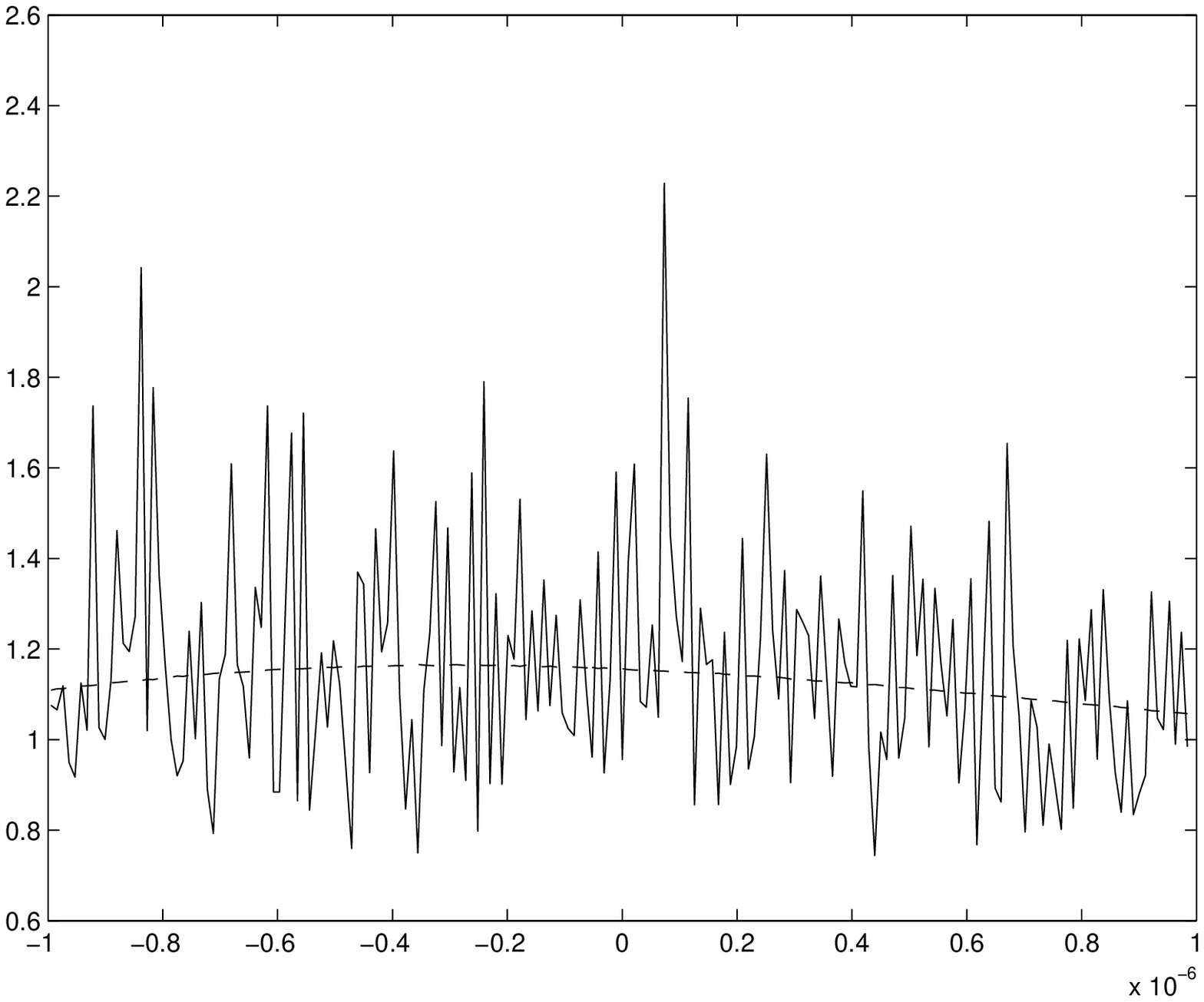}
\put(-40,-10){$y$} \put(-350,220){$E_q^f$} \caption{\label{fluxflux}Fractional error in
the heat flux for ``Temperature Couette'' in a channel
as a function of the transverse channel coordinate, $y$. The
dashed line denotes equation (\ref{heatfluxfluct}) and the solid line
indicates DSMC simulation results. }
\end{center}
\end{figure}

The expressions for shear stress and heat flux were verified using
Couette (walls at equal temperature with different velocities) and
``temperature'' Couette (walls at zero velocity with different
temperatures) calculations respectively. In these calculations,
very small cell sizes ($\Delta x = \lambda/6$) and time steps
$(\Delta t =\lambda /(30 c_\mathrm{m}))$ were used in order to
minimize the discrepancy between the volume-averaged and surface-averaged shear stress and heat flux \cite{frankcell}. The system
was equilibrated for $10^6$ time steps and samples were taken
every 50 time steps. The momentum  and energy fluxes de-correlate faster than the
conserved hydrodynamic variables, such as density, so independent
samples are obtained after fewer time steps (see section~4). Good
agreement is found between the theoretical results and simulation
measurements for volume averaged and fluxal quantities, as shown
in figures \ref{volumestress}, \ref{fluxstress}, \ref{volumeflux},
\ref{fluxflux}.

A final note: In DSMC simulations one considers each particle as
``representing'' a large number of molecules in the physical
system. In all the expressions given above, $N_0$ and $N^+$ relates to  the number of
particles used by the simulation so the fluctuations can be
reduced by using larger numbers of particles (i.e., using a lower
molecule-to-particle ratio).

\subsection{Dense fluids}
We performed molecular dynamics simulations to test the validity
of equations (\ref{velfluct}), (\ref{densfluct}), (\ref{tempfluct}) for
dense fluids. A similar geometry to the dilute gas simulations
described above was used but at a significantly higher density. In
particular, we simulated liquid argon ($\sigma_{LJ}= 3.4\times
10^{-10}$m, $\varepsilon_{LJ}=119.8k_b$) at $T=240K$ and
$\rho=860Kg/\mathrm{m}^3$ in a two-dimensional channel with the
$x$ and $z$ directions periodic. The channel height was
$H=69.7\sigma_{LJ}$. The wall molecules were connected to fcc
lattice sites through springs and interacted with the fluid through
a Lennard-Jones potential with the same parameters. The spring
constant $k_s=460\varepsilon\sigma^{-2}$ was chosen in the way that
root mean square displacement of wall atoms around their
equilibrium position at the simulated temperature was well below
the Lindermann criterion for the melting point of a solid. The
length and depth of the system was $28\sigma_{LJ}$ and
$29.1\sigma_{LJ}$ in the $x$ and $z$ directions respectively. A
constant force $f= 8 \times 10^{-5} \varepsilon/\sigma_{LJ}$ per
particle was used to generate a velocity field with a maximum
velocity of approximately $13$~m/s.

In order to calculate the fluctuation of density, temperature and
velocity, we divided the simulation cell into 13 layers in the
$y-$direction with a height $\Delta y= 4.6396\sigma$. We further
divided each layer into $49$ cells, 7 in each of the $x$ and $z$
directions. The density, temperature and velocity in each cell
were calculated in every $2000(0.005t_{LJ})$, where
$t_{LJ}=\sqrt{m_{LJ}\sigma_{LJ}^2/\varepsilon_{LJ}}$. We have
checked that this time interval is  longer than the system's
correlation time such that samples taken between such intervals
are independent. For each cell, $200$ samples are used to
calculate the average density, temperature and velocity. The
fluctuation was calculated for each layer using the 49 equivalent
cells in the $x-z$ plane.

\begin{figure}
\begin{center}
\includegraphics[width=4.1in]{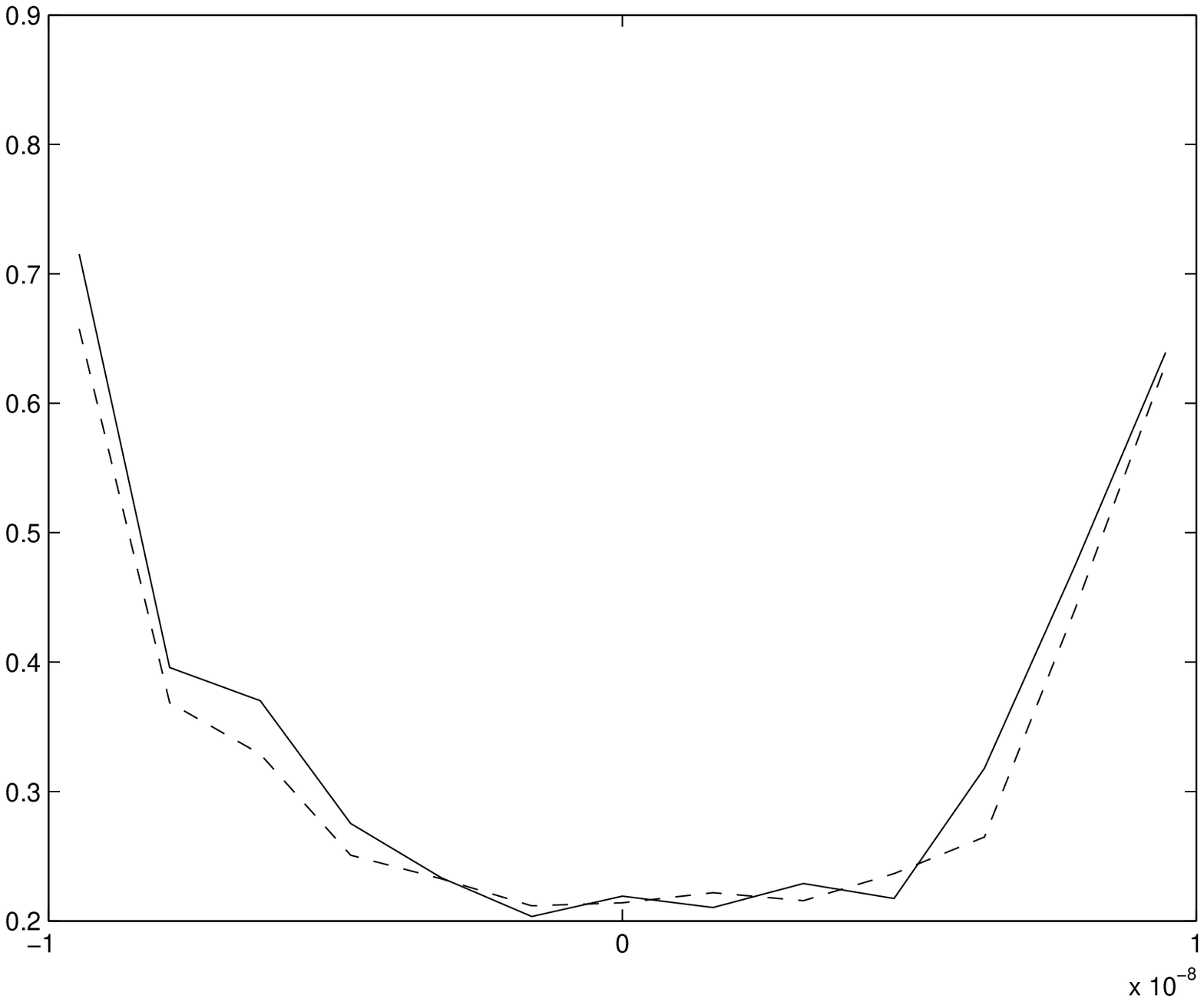}
\put(-40,-10){$y({ \mathrm m})$} \put(-350,220){$E_u$} \caption{\label{mdvelfluct}Fractional error in velocity for dense-fluid Poiseuille flow in a channel as a
function of the transverse channel coordinate, $y$. The dashed
line denotes equation (\ref{velfluct}) and the solid line denotes
MD simulation results.}
\end{center}
\end{figure}

\begin{figure}
\begin{center}
\includegraphics[width=4.1in]{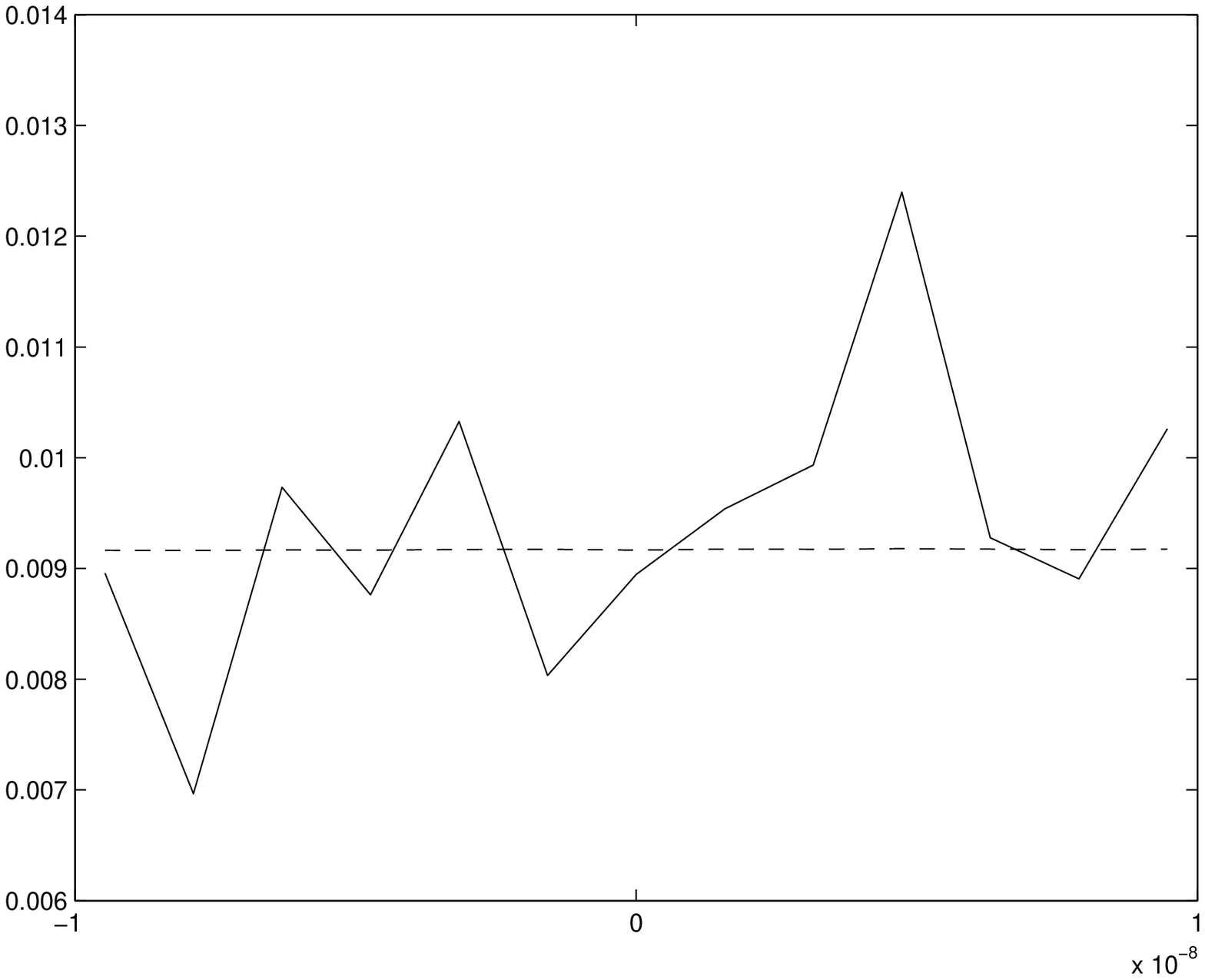}
\put(-40,-10){$y({ \mathrm m})$} \put(-350,220){$E_T$} \caption{\label{mdtempfluct}Fractional error in temperature for dense-fluid Poiseuille flow in a channel
as a function of the transverse channel coordinate, $y$. The
dashed line denotes equation (\ref{tempfluct}) and the solid line
indicates MD simulation results.}
\end{center}
\end{figure}

\begin{figure}
\begin{center}
\includegraphics[width=4.1in]{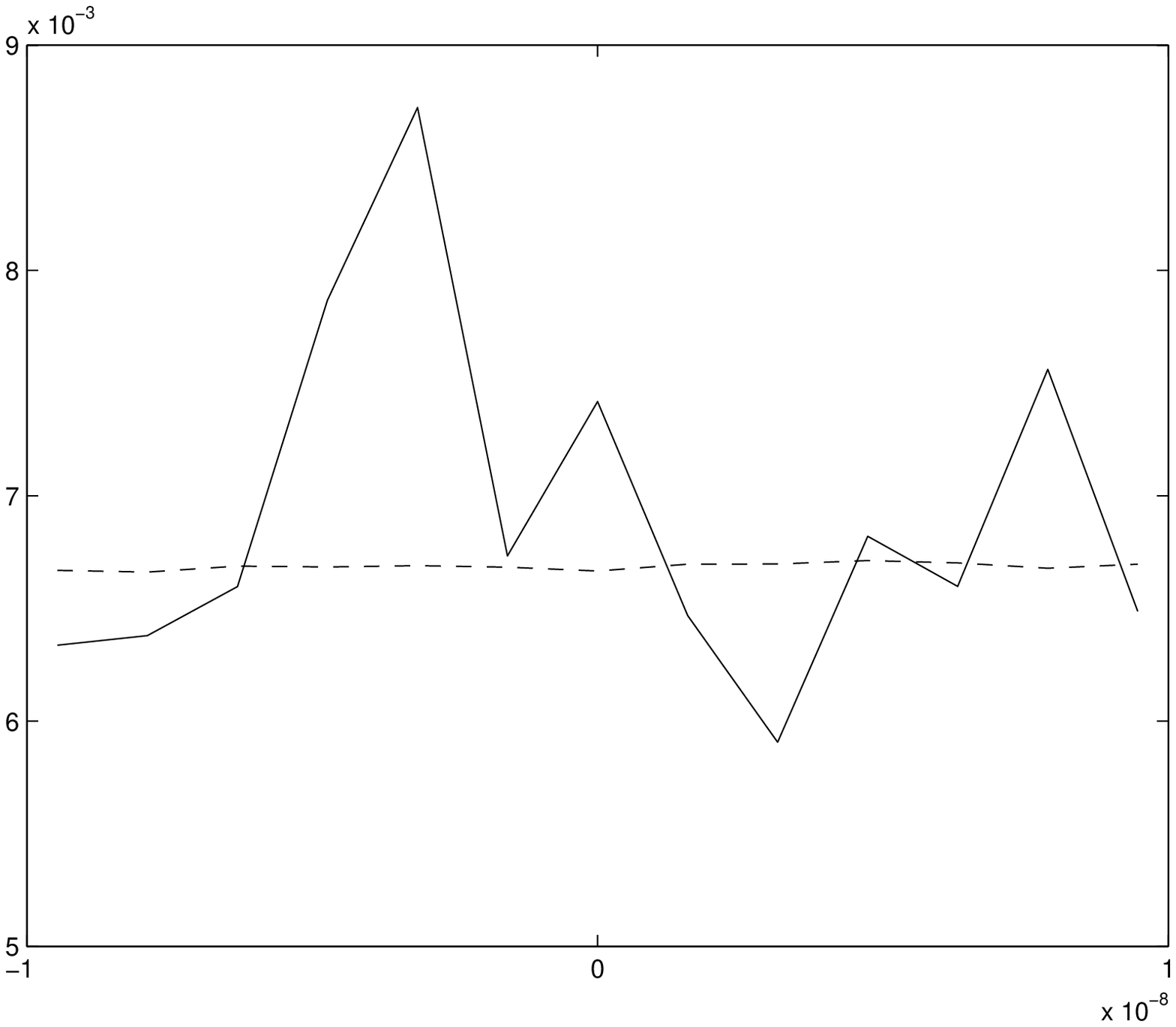}
\put(-40,-10){$y({ \mathrm m})$} \put(-350,220){$E_{\rho}$} \caption{\label{mddensfluct}Fractional error in density for dense-fluid Poiseuille flow in a
channel as a function of the transverse channel coordinate, $y$.
The dashed line denotes equation (\ref{densfluct}) and the solid
line indicates MD simulation results.}
\end{center}
\end{figure}

Due to the sensitivity of the compressibility $\kappa_{T}$ on the
interaction cutoff $r_c$, a rather conservative value of
$r_c=4.0\sigma_{LJ}$ was used. We also introduced a correction for
the still-finite cutoff which used the compressibility predictions
of the Modified Benedict-Webb-Rubin equation of state
\cite{LJEOS}. The agreement between the theoretical predictions
and the simulations is good (see Figures \ref{mdvelfluct}, \ref{mdtempfluct} and \ref{mddensfluct}).

\section{Independent Samples and Correlations}
The results in figures \ref{volumestress}, \ref{fluxstress},
\ref{volumeflux}, \ref{fluxflux} suggest that volume-averaged
measurements provide a superior performance due to a smaller
relative error. This conclusion, however, is not necessarily
correct because our results are based on arbitrary choices of
``measurement spacing'', in the sense that the only consideration
was to eliminate correlations in the data since the theoretical
formulation in section \ref{theorysection} is based on the
assumption of uncorrelated samples. The two methods of sampling
are in fact linked by a very interesting interplay between the
roles of time and space: fluxal sampling is a measurement at a
{\it fixed} position in space for a period of time, whereas volume
sampling is performed over some region of space at a {\it fixed}
time. The theoretical performance of each method can be increased
by extending the respective window of observation. However, by
increasing the time of observation, a fluxal measurement becomes
correlated with neighboring measurements if the same particle
crosses more than one measuring station within the period of
observation. Similarly, by increasing the region of measurement,
 subsequent volume measurements will suffer from
time correlations if previously interrogated particles do not have
sufficient time to leave the measurement volume. This relation
between spatial and temporal sampling and the role of the particle
characteristic velocity is also manifested in the theoretical
predictions. Enforcing equality of the variances of the respective volume and
fluxal measurements (eqs (\ref{tauVarEqn2}) and
(\ref{stressfluct}), and eqs (\ref{qVarEqn2}) and
(\ref{heatfluct})) yields $\Delta x =\beta c_\mathrm{m} \Delta t$
where $\beta \approx 1$. The generalization of this work to
include time and spatial correlations is the subject of future
work.

The effect of time correlation in volume measurements can be
approximated using the theory of the ``persistent random
walk''~\cite{weiss}, first introduced by F\"urth~\cite{furth} and
Taylor \cite{taylor}.  A persistent random walk is one in which each
step displacement, $u(t_\eta) = u(\eta \Delta t)$, is identically
distributed and has a positive correlation coefficient with the
previous step,
\begin{equation}
\alpha = \frac{\langle \delta u(t_\eta) \; \delta
u(t_{\eta-1}) \rangle}{\langle \delta u^2 \rangle} ,  \label{alpha}
\end{equation}
($0 < \alpha < 1$), e.g. to model diffusion in a turbulent
fluid~\cite{taylor,ghosal}. This assumption implies that step
correlations decay exponentially in time,
\begin{equation}
\frac{\langle \delta u(t_\eta) \; \delta u(0) \rangle}{\langle \delta u^2
\rangle} = \alpha^\eta = e^{-t_\eta/t_c}, \label{corr}
\end{equation}
where $t_c = -\Delta t / \log \alpha$ is the correlation time, beyond
which the steps are essentially indepedent.  The position, $U(t_\eta)$,
of the random walker after $\eta$ steps is the sum of these correlated
random displacements, $U(t_\eta) = \sum_{j=1}^\eta u(t_j)$. Following
Taylor, it is straightforward to show that for times long compared to the correlation time ($t_\eta \gg t_c$), the usual diffusive scaling holds
\begin{equation}
\langle \delta U(t_\eta)^2 \rangle \sim \langle \delta u^2 \rangle\; \eta \;
\left(\frac{1+\alpha}{1-\alpha}\right),
\end{equation}
where the bare diffusion coefficient is modified by the term in
parentheses due to correlations.

There is a natural connection with the measurement of statistical
averages. If we view each step in the persistent random walk as a
correlated sample of some quantity in the gas, then the position of
the walker (divided by the number of samples) corresponds to the
sample average. Thus the variance of a set of sequentially correlated random
variables $\{u(t_\eta)\}$, $Var^c(u)$, may  be written as
\begin{equation}
Var^c(u) = Var(u) \frac{1 + \alpha}{1 - \alpha}
\end{equation}
where $Var(u)$ is the variance of the uncorrelated samples ($\alpha=0$). 
The theory above implies that the sample variance is
amplified by the presence of correlations, 
because effectively fewer independent samples have been taken compared
to the uncorrelated case, $\alpha = 0$.

Note that a sequence of correlated random variables, $\{ u(t_\eta)\}$,
satisfying Eqs.~(\ref{alpha}) and (\ref{corr}) can be explicitly
constructed from a sequence of independent, identically distributed
variables, $\{ \tilde{u}(t_\eta)\}$, by letting $u(t_\eta)$ equal the
previous value $u(t_{\eta-1})$ with probability $\alpha$ or a new value
$\tilde{u}(t_\eta)$ with probability $1-\alpha$. This allows us to
interpret $\alpha$ as the probability that a sample is the same as the
previous one, which is precisely the source of correlations when 
volume-sampling dilute gases.

There are two distinct ways that a new sample of a dilute gas can
actually provide new information: (i) Either some new particles have
entered the sampling cell, or (ii) some particles previously
inside the cell have changed their properties due to
collisions.  Regarding (i), the probability that a particle will
remain in a cell of size $\Delta x$ after time step $\Delta t$ is
given by
\begin{equation}
\alpha_l=\left(\frac{1}{\Delta x}\int_{-\Delta x/2}^{\Delta x/2}
\int_{-\Delta x/2}^{\Delta x/2} f\left(\frac{x-\tilde{x}}{\Delta t}\right) dx d\tilde{x}\right)^d
\label{escape}
\end{equation}
where $d$ is the dimensionality of the cell and $f(c_i)$ is the
probability distribution function of the particle velocity, and thus
$f(\frac{x-\tilde{x}}{\Delta t})$ represents the probability that a
particle originating from location $\tilde{x}$ is found at $x$ after a
time interval, $\Delta t$. Note that the above expression holds when
the components of the particle velocity in different directions are
uncorrelated and that the particle spatial distribution inside the
cell is initially uniform.  Regarding (ii), if the single-particle
auto-correlation function decays exponentially, the ``renewal''
probability $1-\alpha_c$ for collisional de-correlation can, at least
in principle, be inferred from simulations using Eq.~(\ref{corr}).
The net correlation coefficient, giving the probability of an
``identical'' sample, is $\alpha=\alpha_c\,\alpha_l$.

\begin{figure}
\begin{center}
\includegraphics[width=4.1in]{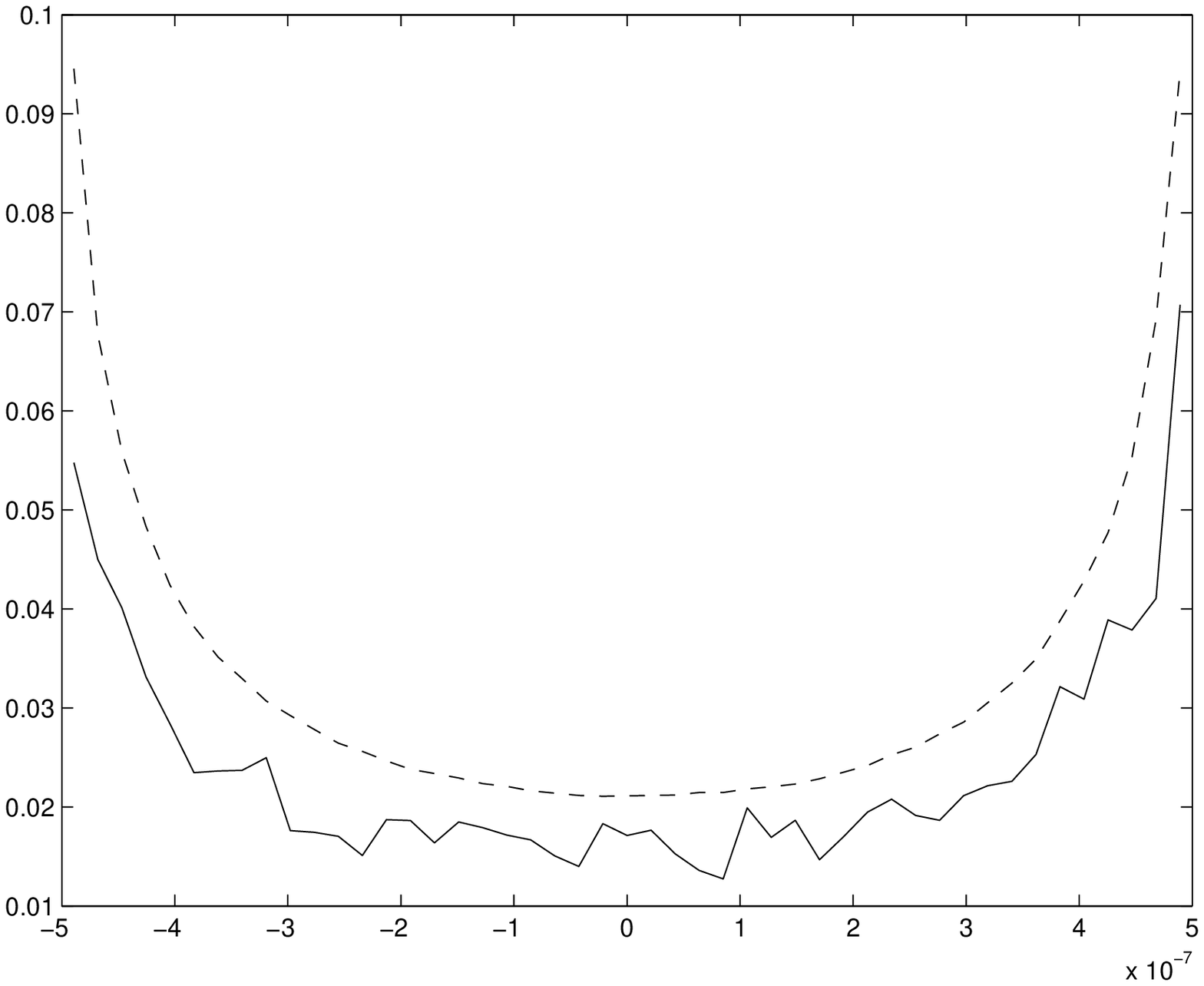}
\put(-40,-10){$y$} \put(-350,220){$E_u$} \caption{ \label{velocityfigcorr}
Fractional error in velocity for Poiseuille flow in a channel as a
function of the transverse channel coordinate, $y$. The dashed
line denotes the theoretical prediction including correlations and the solid line denotes
DSMC simulation results.}
\end{center}
\end{figure}

\begin{figure}
\begin{center}
\includegraphics[width=4in]{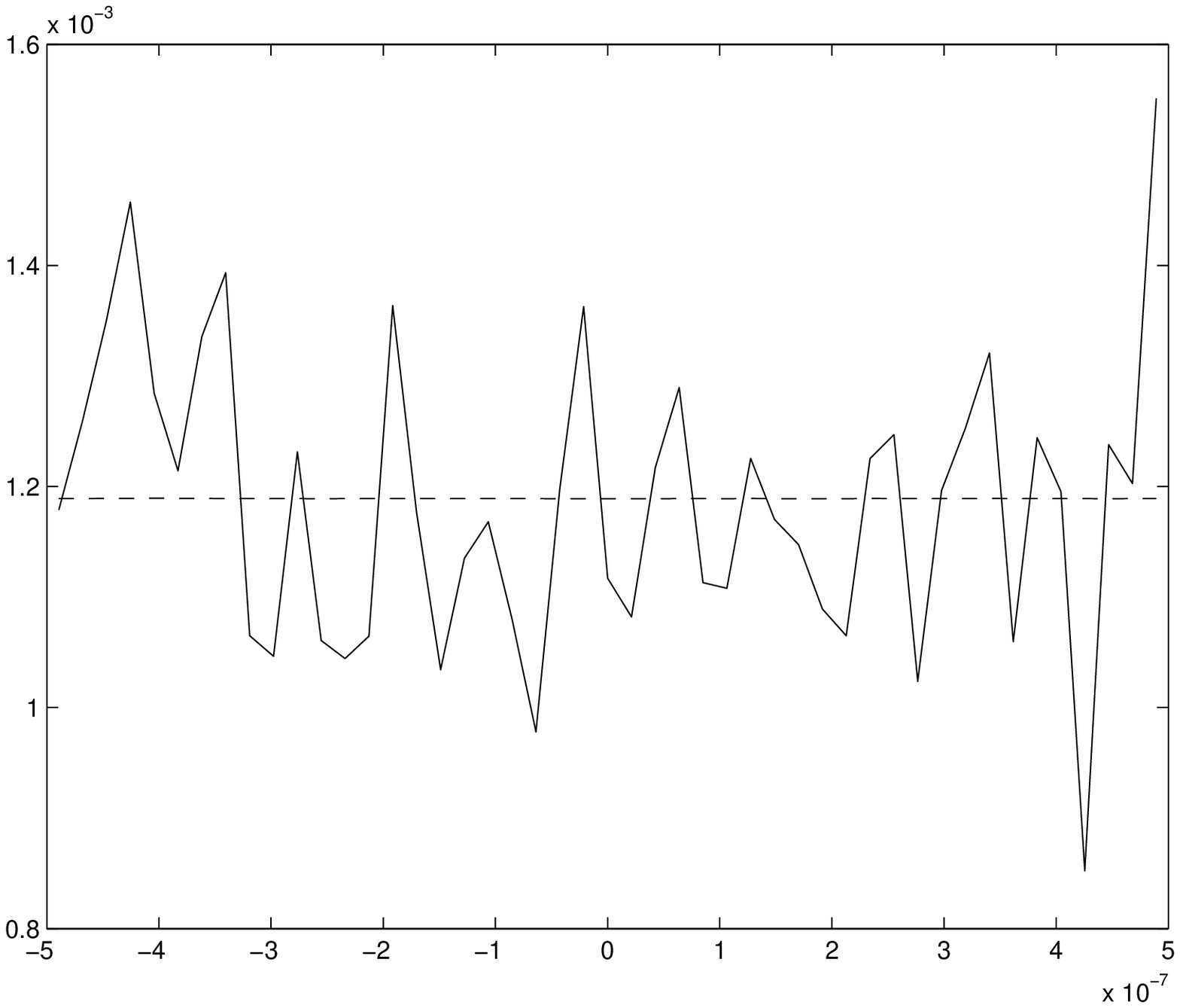}
\put(-40,-10){$y$} \put(-350,220){$E_{\rho}$} \caption{
\label{densityfigcorr} Fractional error in density for Poiseuille flow in a
channel as a function of the transverse channel coordinate, $y$.
The dashed line denotes the theoretical prediction including correlations and the solid line indicates DSMC simulation results.}
\end{center}
\end{figure}

\begin{figure}
\begin{center}
\includegraphics[width=4in]{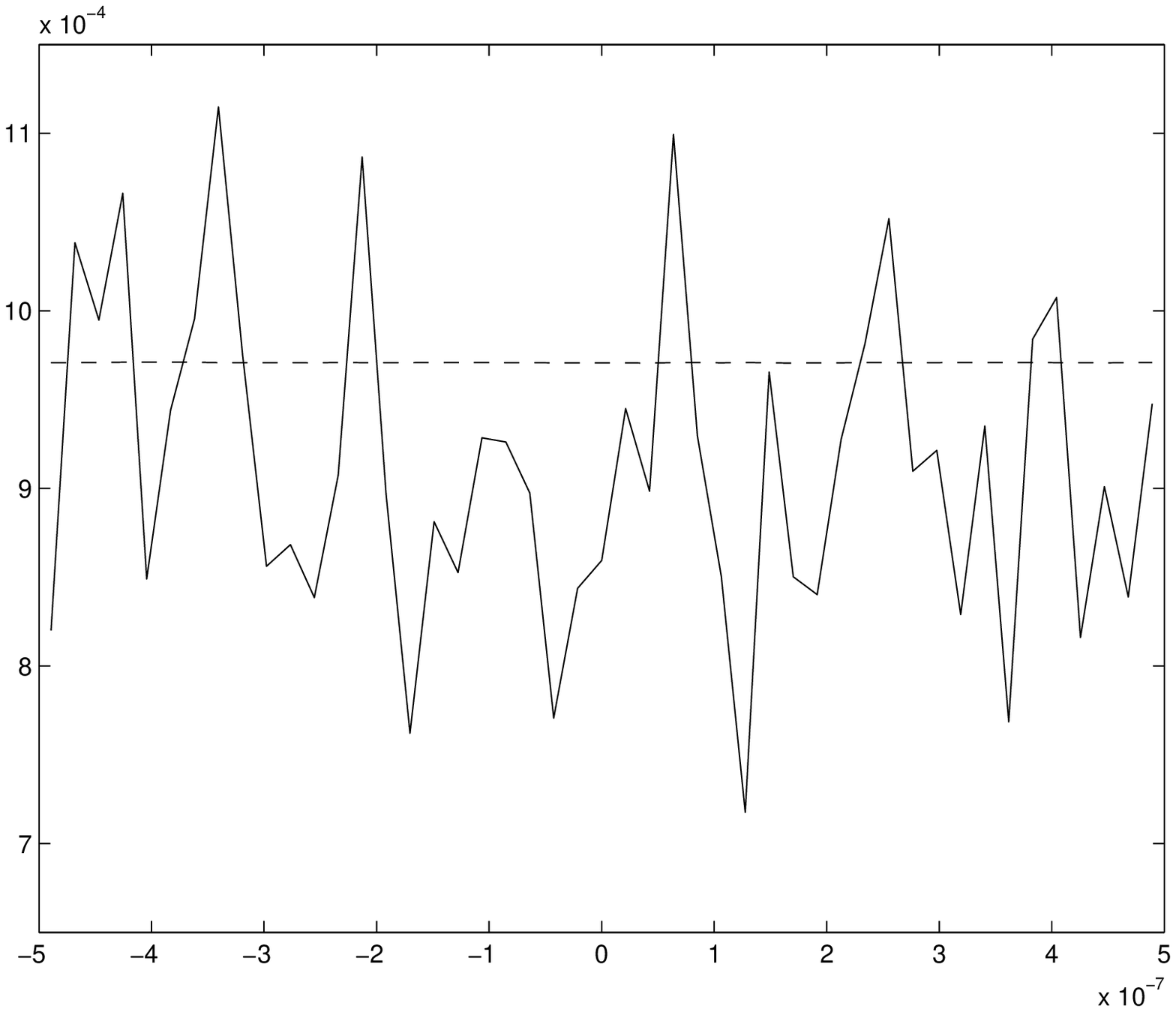}
\put(-40,-10){$y$} \put(-350,220){$E_T$} \caption{ \label{temperaturefigcorr}
Fractional error in temperature for Poiseuille flow in a channel
as a function of the transverse channel coordinate, $y$. The
dashed line denotes the theoretical prediction including correlations and the solid line
indicates DSMC simulation results.}
\end{center}
\end{figure}

In our DSMC simulations, the single-particle correlation coefficient
for the velocity was estimated to be $\alpha_c\approx 0.94 $ for
$\Delta t= \lambda/(15c_\mathrm{m})$ by fitting Eq.~(\ref{corr})
to data; the same value for $\alpha_c$ was found in both equilibrium
and non-equilibrium simulations.  In the comparison of figures
\ref{velocityfigcorr}, \ref{densityfigcorr} and
\ref{temperaturefigcorr}, however, we use $\alpha_c=1$ since mass,
momentum and energy are conserved during collisions. We find that this
analysis produces very good results in the case of density and
temperature (see figures \ref{densityfigcorr} and
\ref{temperaturefigcorr}) and acceptable results for the case of mean
velocity (see figure \ref{velocityfigcorr}). Use of $\alpha_c=0.94$
would tend to make the agreement better but no theoretical
justification exists for it. The value of $\alpha_l$ was directly
calculated by assuming an equilibrium distribution. For our
two-dimensional calculations with $\Delta x/(c_\mathrm{m}\Delta
t)=2.5$, we find $\alpha_l\approx 0.6$. The effect of a mean flow is
very small if the Mach number is small. This was verified through both
direct evaluation of Eq.~(\ref{escape}) using a local equilibrium
distribution function and DSMC simulations.

\section{Conclusions}

We have presented expressions for the statistical error in
estimating the velocity, density, temperature and pressure in
molecular simulations. These expressions were validated for flow
of a dilute gas and dense liquid in a two-dimensional channel
using the direct simulation Monte Carlo and Molecular Dynamics
respectively. Despite the non-equilibrium nature of the validation
experiments, good agreement is found between theory and
simulation, verifying that modifications to non-equilibrium
results are very small. In particular, in the dense fluid case,
despite the significant non-equilibrium due to a shear of the
order of $5.5\times 10^8~\mathrm {s}^{-1}$, the agreement with
equilibrium theory is remarkable. We thus expect these results to
hold for general non-equilibrium applications of interest.

Predictions were also presented for the statistical error in
estimating the shear stress and heat flux in dilute gases through
cell averaging and surface averaging. Comparison with direct Monte
Carlo simulations shows that the equilibrium assumption is
justified. Within the same sets of assumptions, we were able to
show that the distribution of particles leaving a cell is
Poisson-distributed.

One consideration that significantly limits the applicability of
cell (volume) averaging for transport quantities is their neglect
of transport due to collisions in the cell volume. In DSMC in
particular, as the time step of the simulation is increased, the
fraction of particles in a cell undergoing collision increases,
and as a result, the error from using the above method increases.

It was found that the fluctuation in state variables is
significantly smaller compared to flux variables in the continuum
regime $Kn \rightarrow 0$. This is important for the development
of hybrid methods. Although a direct comparison was only presented
between volume averaged quantities, we find  that the fluxal
measurements for the shear stress and heat flux perform similarly
to the volume averaged counterparts (regarding scaling with the
Knudsen number).

\section*{Appendix: Number Fluctuations in Dilute Gases}

Consider an infinite ideal gas in equilibrium with mean number
density, $n$. By definition, each infinitessimal volume element,
$dV$, contains a particle with probability, $n dV$, and each such
event in independent. From these assumptions, it is straightforward to
show that the number of particles $N$ in an arbitrary volume $V$ is a
Poisson random variable~\cite{feller},
$$
\mbox{Prob}(N=m) =  \frac{  e^{-\langle N \rangle} {\langle N
\rangle}^m }{m!}
$$
with mean and variance given by
$$
\langle N \rangle = \langle \delta N^2 \rangle = n V.
$$
Note that this result does not strictly apply to a finite ideal gas
because the probabilities of finding particles in different
infinitessimal volumes are no longer independent, due to the global
constraint of a fixed total number of particles. Nevertheless, it is
an excellent approximation for most dilute gases, even finite
non-ideal gases far from equilibrium (e.g. as demonstrated by
simulations in the main text).

Now consider a property $A$ which each particle in a volume $V$ may
possess independently with probability, $P_A$.  In this Appendix, we
show that the distribution of the number of such particles, $N_A$, is
also a Poisson random variable with mean and variance given by
$$
\langle N_A \rangle = \langle \delta N_A^2 \rangle = P_A \langle N
\rangle = P_A n V.
$$
For example, in the main text we require the number of particles,
$N_{ij}$, which travel from one region, $i$, to another region, $j$,
in a time interval, $dt$. The proof given here, however, is much more
general and applies to arbitrary number fluctuations of an infinite
ideal gas in equilibrium, such as the number of particles in a certain
region, moving in a certain direction, of a certain ``color'', with
speeds above a certain theshold, etc.

We begin by expressing $N_A$ as a random sum of random variables,
\begin{equation}
N_A=\sum^N_{i=1}\chi_i
\label{sum}
\end{equation}
where
\[
\chi_i=\left\{
\begin{array}
{r@{\quad }l}
1 & \mathrm{if}\; A\; \mathrm{occurs}\\
0 & \mathrm{otherwise.}
\end{array}
\right.
\]
is an indicator function for particle $i$ to possess property $A$,
which is a Bernoulli random variable with mean, $P_A$. It is
convenient to introduce probability generating functions,
$$f_\chi(z)=\sum^\infty_{m=0}\mbox{Prob}(\chi=m)z^m = (1-P_A) + P_A z$$
and
\begin{equation}
f_N(z)=\sum^\infty_{m=0}\mbox{Prob}(N=m)z^m=
\sum^\infty_{m=0} \frac{ e^{-\langle N \rangle} {\langle N
\rangle}^m z^m}{m!}=e^{\langle N \rangle (z-1)}
\label{poissongen}
\end{equation}
because the generating function for a random sum of random variables,
as in Eq.~(\ref{sum}), is simply given by a composition of the
generating functions for the summand and the number of
terms~\cite{feller},
$$f_{N_A}(z)=\sum_{m=0}^\infty \mbox{Prob}(N_A =m)z^m =
f_N\left(f_\chi (z)\right).$$
Combining these expressions we have
$$f_{N_A} (z)=e^{\langle N \rangle(1-P_A+P_Az-1)} =e^{P_A\langle N
\rangle(z-1)}.$$ Comparing with Eq.~(\ref{poissongen}) completes the
proof that $N_A$ is a Poisson random variable with mean, $P_A \langle
N \rangle$.

\section{Acknowledgements}
The authors wish to thank M. Malek-Mansour and B. Alder for
helpful discussions. The authors would also like to thank X.
Garaizar for making this work possible through the computer
resources made available to them. This work was also supported, in
part, by a grant from the University of Singapore, through the
Singapore-MIT alliance.

\end{document}